# Integrals of motion of classical lattice sine-Gordon system

B. Enriquez and B.L. Feigin

**Abstract.** *We compute the local integrals of motions of the classical limit of the lattice sine-Gordon system, using a geometrical interpretation of the local sine-Gordon variables. Using an analogous description of the screened local variables, we show that these integrals are in involution. We present some remarks on relations with the situation at roots of 1 and results on another latticisation (linked to the principal subalgebra of $\widehat{s\ell}_2$ rather than the homogeneous one). Finally, we analyse a module of "screened semilocal variables", on which the whole $\widehat{s\ell}_2$ acts.*

## Introduction.

In this paper, we analyse the classical limit of a lattice version of the sine-Gordon system. It consists of $q$-commuting variables set on a line lattice, representing as usual densities for the screening charges.

We first determine expressions for the local integrals of motions of this system. For this, we formulate this problem in terms of the cohomology of the action of the screening operators on the module generated by a finite number of lattice variables. It has been known for some time ([F]) that this is an action of the nilpotent part of $\widehat{s\ell}_2$; here we interpret the space of lattice variables as an homogeneous space for this action. We find that lattice variables are coordinates on the Schubert cells of $SL_2(\mathbf{C}((\lambda^{-1})))$, which are an affine version of the Demazure desingularisation. We give explicit formulae for the cocyles and the integrals of motion.

We then solve the same problem (less explicitly) for an other lattice version of sine-Gordon, which can be generalised only to $\widehat{s\ell}_n$ (whereas in our main approach the problem can be formulated for arbitrary affine Kac-Moody algebras).

Coming back to the main setting, we show that the integrals of motion commute in Poisson sense. For this, we study their action on the variables on the whole line. To this end, we analyze first "screened" variables on the line (that is, the module generated by the screening action on the lattice variables); they are endowed with an action of the whole algebra $\widehat{s\ell}_2$ (at level zero). We are then able to show that the Hamiltonian vector fields generated by the integrals of motion correspond to the action of a commutative Lie algebra on a homogeneous space of $\widehat{s\ell}_2$.

We then study "screened semilocal quantities", containing in addition to the variables above "half integrals of motion" (that is, integrals on the half line of densities of integrals of motion). We still obtain a homogeneous space of $\widehat{s\ell}_2$, equal (up to completion) to $SL_2(\mathbf{C}((\lambda^{-1})))/H$ (where $H$ is the Cartan subgroup). We study the Poisson structure of this space; it is connected to the trigonometric $r$-matrix. This study enables us to precise the result on the Hamiltonian vector fields generated by the integrals of motion.

We hope that this group theoretic interpretation will be useful for constructing solitonic solutions of this system.

Let us note finally that related lattice approaches to the sine-Gordon systems were developed by Izergin and Korepin ([IK1], [IK2]).

## 1. Formulation of the problem of integrals of motion.

Consider a system of variables $x_i, y_i, i \in \mathbf{Z}$, with relations $x_i x_j = q x_j x_i, y_i y_j = q y_j y_i, x_i y_j = q^{-1} y_j x_i, y_i x_j = q^{-1} x_j y_i$ for $i < j$, and $x_i y_i = q^{-1} y_i x_i$ (we can consider this system as obtained from $z_i z_j = q z_j z_i, i < j$ by posing $x_i = z_{2i}, y_i = z_{2i+1}^{-1}$). The operators $\Sigma^+ = \sum_{i \in \mathbf{Z}} x_i$ and $\Sigma^- = \sum_{i \in \mathbf{Z}} y_i$ are lattice analogues of the screening charges of the $\widehat{sl}_2$ Toda system. Lattice analogues of local conservation laws are quantities $\sum_{i \in \mathbf{Z}} P(x_i, y_i, \cdots, x_{i+d})$, commuting with $\Sigma^+$ and $\Sigma^-$. We are going to determine the classical limits of these conservation laws.

## 2. Cohomological interpretation and classical limit.

Let us give variables $x_i$ and $y_i$ the degrees $+1$ and $-1$ and pose $[a,b]_q = ab - q^{\deg a \deg b} ba$. Then there is an action of the negative part $U_q \widehat{b}_-$ of $U_q \widehat{sl}_2$ on $\mathbf{C}[x_i, y_i, \bar{\Sigma}^+, \bar{\Sigma}^-]$ by $\widehat{Q}_\pm(x) = [\Sigma^\pm, x]_q, \widehat{K}(x) = q^{\deg x} x$ (here $\bar{\Sigma}^+ = \sum_{i=-1}^{-\infty} x_i$ and $\bar{\Sigma}^- = \sum_{i=-1}^{-\infty} y_i$). The polynomial $P$ will be a density for a conservation law if there exist polynomials $\psi^\pm$ such that $\widehat{Q}_\pm P = (T-1) \psi^\pm$ ($T$ is the shift operator). Then $\psi^\pm$ will be a 1-cocycle for the action of $U_q \widehat{b}_-$ in $\mathbf{C}[x_i, y_i, \bar{\Sigma}^\pm]$. So the problem of quantum conservation laws is equivalent to the computation of cohomology classes $\psi \in H^1(U_q \widehat{b}_+, \mathbf{C}[x_i, y_i, \bar{\Sigma}^\pm])$, such that $\psi(\widehat{Q}^\pm)$ are independent of $\bar{\Sigma}^\pm$, and of the map $T-1$ in cohomology.

We will be interested in the classical version of this problem. The classical limits of operators $\widehat{Q}_\pm, \widehat{K}$ form an action of $\widehat{b}_-$ by vector fields

$$\bar{Q}_+ = \sum_{i \in \mathbf{Z}} e^{\tau_i} \left( \frac{1}{2} \frac{\partial}{\partial \tau_i} + \sum_{j > i} \frac{\partial}{\partial \tau_j} - \sum_{j \geq i} \frac{\partial}{\partial \sigma_j} \right), \quad \bar{Q}_- = \sum_{i \in \mathbf{Z}} e^{\sigma_i} \left( \frac{1}{2} \frac{\partial}{\partial \sigma_i} + \sum_{j > i} \frac{\partial}{\partial \sigma_j} - \sum_{j > i} \frac{\partial}{\partial \tau_j} \right),$$

$$(*) \qquad \bar{H} = 2 \sum_{i \in \mathbf{Z}} \left( \frac{\partial}{\partial \tau_i} - \frac{\partial}{\partial \sigma_i} \right),$$

where $x_i = e^{\tau_i}$ and $y_i = e^{\sigma_i}$.

Let us suppose that $\psi^\pm$ depend on variables $x_0, y_0, \cdots, x_k, y_k$ ; we can write the operators $\bar{Q}_\pm$ on $\mathbf{C}[x_0, \cdots, y_k] \otimes \mathbf{C}[\bar{\Sigma}_\pm, \{\bar{\Sigma}_+, \bar{\Sigma}_-\}', \cdots]$ as $\bar{Q}_\pm = Q_\pm \otimes 1 \pm \frac{1}{2} H \otimes m(\bar{\Sigma}_\pm) + 1 \otimes \{\bar{\Sigma}_\pm, \cdot\}'$, where $m(\bar{\Sigma}_\pm)$ is the operator of multiplication by $\bar{\Sigma}_\pm$, and $\{a,b\}' = \{a,b\} - (\deg a)(\deg b) ab$. Here $Q_\pm, H$ are vector fields on $\mathbf{C}[x_0, \cdots, y_k]$ given by formulae $(*)$ with summation over $i = 0, \cdots, k$ ; they still form an action of $\widehat{b}_-$.

**Lemma.—** *The polynomials $\psi^\pm$ form a cocycle (resp. coboundary) for the action of $\widehat{n}_-$ by operators $\bar{Q}_\pm$ on $\mathbf{C}[x_0, \cdots, y_k] \otimes \mathbf{C}[\bar{\Sigma}_\pm, \cdots]$, if an only if they form a cocycle (resp. coboundary) for the action of $\widehat{n}_-$ by $Q_\pm$ on $\mathbf{C}[x_0, \cdots, y_k]$.*

**Proof.** The coboundary statements are obvious. For the cocycle statements, the first implies the second by consideration of terms of type $x \otimes 1$ in the identities. That the second statement implies the first is the result of a long but straightforward computation. ∎

Note that the condition that $\deg \psi^+ = -\deg \psi^- = 1$ is equivalent to the condition that the cocycle $\psi^\pm$ can be prolongated to a cocycle of $\widehat{b}_-$, with $\psi(H) = a$ constant.

### 3. Homogeneous spaces under $\widehat{b}_-$

Recall that the representation of $U_q \widehat{n}_-$ in $q$-commuting variables comes from the sequence of morphisms $U_q \widehat{n}_- \xrightarrow{\Delta^{2k}} (U_q \widehat{n}_-)^{\overline{\otimes} 2k} \to \mathbf{C}[x_i, y_i]_{i=1,\cdots,2k}$, where the last map is the alternating product of morphisms $U_q \widehat{n}_- \to \mathbf{C}[x_i], Q_+ \mapsto x_i, Q_- \mapsto 1$, and $U_q \widehat{n}_- \to \mathbf{C}[y_i], Q_+ \mapsto 1, Q_- \mapsto y_i$. Recalling the Hopf algebra isomorphism $U_q \widehat{n}_- \simeq \mathbf{C}[\widehat{N}_+]_q$, we see that the map $U_q \widehat{n}_- \to \mathbf{C}[x_i, y_i]$ has for classical limit the embedding of $\mathbf{C}^2$ in $\widehat{N}_+$, $(x_i, y_i) \mapsto \begin{pmatrix} 1 & 0 \\ \lambda x_i & 1 \end{pmatrix} \begin{pmatrix} 1 & y_i \\ 0 & 1 \end{pmatrix}$. Accordingly, the classical limit of $\mathbf{C}[\widehat{N}_+]_q \simeq U_q \widehat{n}_- \to \mathbf{C}[x_i, y_i]_{i=1,\cdots,k}$ is dual to the map $\mathbf{C}^{2k} \to \widehat{N}_+, (x_i, y_i) \mapsto \prod_{i=1}^{k} \begin{pmatrix} 1 & 0 \\ \lambda x_i & 1 \end{pmatrix} \begin{pmatrix} 1 & y_i \\ 0 & 1 \end{pmatrix}$. More generally, we have also algebra morphisms $\mathbf{C}[\widehat{N}_+]_q \to \mathbf{C}[x_1, y_1 \cdots, x_N, y_N, x_{N+1}]$ whose classical limit are dual to $(x_1, y_1 \cdots x_{N+1}) \mapsto \begin{pmatrix} 1 & 0 \\ \lambda x_1 & 1 \end{pmatrix} \begin{pmatrix} 1 & y_1 \\ 0 & 1 \end{pmatrix} \cdots \begin{pmatrix} 1 & 0 \\ \lambda x_{N+1} & 1 \end{pmatrix}$, etc. These maps $\mathbf{C}^N \to \widehat{N}_+$ are Poisson morphisms, in fact the image of $(\mathbf{C}^*)^N$ is dense in a symplectic leaf of $\widehat{N}_+$. Indeed, images of $\mathbf{C}^N$ have dimension $N$, and images of $N$-uples have the respective forms

- for $(x_1, y_1 \cdots x_k, y_k)$, $\begin{pmatrix} *\lambda^{k-1} + \cdots & *\lambda^{k-1} + \cdots \\ *\lambda^k + \cdots & *'\lambda^k + \cdots \end{pmatrix}$
- for $(x_1, y_1, \cdots y_{k-1} x_{k+1})$, $\begin{pmatrix} *\lambda^k + \cdots & *\lambda^{k-1} + \cdots \\ *'\lambda^{k+1} + \cdots & *\lambda^k + \cdots \end{pmatrix}$
- for $(y_1, x_2, \cdots y_k, x_{k+1})$, $\begin{pmatrix} *'\lambda^k + \cdots & *\lambda^{k-1} + \cdots \\ *\lambda^k + \cdots & *\lambda^{k-1} + \cdots \end{pmatrix}$
- for $(y_1, x_2, \cdots, x_k, y_k)$, $\begin{pmatrix} *\lambda^{k-1} + \cdots & *'\lambda^{k-1} + \cdots \\ *\lambda^{k-1} + \cdots & *\lambda^{k-1} + \cdots \end{pmatrix}$

(the $*'$ are not zero if the $x_i, y_i$ are not zero), whereas the symplectic leaves of $\widehat{N}_+$ are the preimages of the $\widehat{B}_-$-orbits on $\widehat{G}/\widehat{B}_-$ by the injection $\widehat{N}_+ \to \widehat{G}/\widehat{B}_-$ [here $\widehat{G} = SL_2(\mathbf{C}((\lambda^{-1})))$ and $\widehat{B}_- = \pi^{-1}(B)$, where $\pi : SL_2(\mathbf{C}[[\lambda^{-1}]]) \to SL_2(\mathbf{C})$ is defined by $\lambda^{-1} \mapsto 0$ and $B = \left\{ \begin{pmatrix} t & 0 \\ a & t^{-1} \end{pmatrix} \right\} \subset SL_2(\mathbf{C})$], according to Semenov-Tian-Shansky [STS]. The $\widehat{B}_-$-double cosets in $\widehat{G}$ are the double classes of affine

Weyl group elements, so that the images of $(\mathbf{C}^*)^N$ by these morphisms are dense in $\widehat{B}_-w\widehat{B}_-/\widehat{B}_-$, $w$ respectively equal to

$$\begin{pmatrix} \lambda^{-k} & 0 \\ 0 & \lambda^k \end{pmatrix}, \begin{pmatrix} 0 & -\lambda^{-k-1} \\ \lambda^{k+1} & 0 \end{pmatrix}, \begin{pmatrix} \lambda^k & 0 \\ 0 & \lambda^{-k} \end{pmatrix}, \begin{pmatrix} 0 & -\lambda^{k-1} \\ \lambda^{-k+1} & 0 \end{pmatrix}.$$

For each of these spaces, the vector fields $Q_\pm$ get identified with the natural translation action of $\widehat{b}_-$ on $\widehat{B}_-w\widehat{B}_-/\widehat{B}_-$.

We further identify $\widehat{B}_-w\widehat{B}_-/\widehat{B}_-$ with $\widehat{B}_-/\widehat{B}_- \cap \widehat{B}_-^w$, where $\widehat{B}_-^w = w\widehat{B}_-w^{-1}$; $\widehat{B}_- \cap \widehat{B}_-^w$ are respectively the groups of matrices of the form

$$\begin{pmatrix} a & \lambda^{-2k-1}c \\ b & d \end{pmatrix}, \begin{pmatrix} a & \lambda^{-2k-2}c \\ b & d \end{pmatrix}, \begin{pmatrix} a & \lambda^{-1}c \\ \lambda^{-2k}b & d \end{pmatrix}, \begin{pmatrix} a & \lambda^{-1}c \\ \lambda^{-2k+1}b & d \end{pmatrix}, a,b,c,d \in \mathbf{C}[[\lambda^{-1}]].$$

Finally, $\widehat{B}_-/\widehat{B}_- \cap \widehat{B}_-^w$ can be identified with the varieties $\lambda^{-1}\mathbf{C}[[\lambda^{-1}]]/\lambda^{-2k-1}\mathbf{C}[[\lambda^{-1}]]$, $\lambda^{-1}\mathbf{C}[[\lambda^{-1}]]/\lambda^{-2k-2}\mathbf{C}[[\lambda^{-1}]]$, $\mathbf{C}[[\lambda^{-1}]]/\lambda^{-2k}\mathbf{C}[[\lambda^{-1}]]$, $\mathbf{C}[[\lambda^{-1}]]/\lambda^{-2k+1}\mathbf{C}[[\lambda^{-1}]]$, by associating to $\rho, \rho', \sigma, \sigma'$ in these vector spaces the right cosets of $\begin{pmatrix} 1 & \tilde{\rho} \\ 0 & 1 \end{pmatrix}, \begin{pmatrix} 1 & \tilde{\rho}' \\ 0 & 1 \end{pmatrix}, \begin{pmatrix} 1 & 0 \\ \tilde{\sigma} & 1 \end{pmatrix}, \begin{pmatrix} 1 & 0 \\ \tilde{\sigma}' & 1 \end{pmatrix}$ respectively ($\sim$ is a section of $\lambda^\varepsilon \mathbf{C}[[\lambda^{-1}]] \to \lambda^\varepsilon\mathbf{C}[[\lambda^{-1}]]/\lambda^{-n}\mathbf{C}[[\lambda^{-1}]]$, $\varepsilon = 0$ or $1$). The action of $g = \begin{pmatrix} a & \lambda^{-1}c \\ b & d \end{pmatrix} \in \widehat{B}_-$ on these affine varieties is then given by the homographic transformations

$$g.\rho = \frac{a\rho + c/\lambda}{b\rho + d}, g.\rho' = \frac{a\rho' + c/\lambda}{b\rho' + d}, g.\sigma = \frac{b + d\sigma}{a + c\sigma/\lambda}, g.\sigma' = \frac{b + d\sigma'}{a + c\sigma'/\lambda}.$$

We can make more explicit the maps from $(\mathbf{C}^*)^N$ to $\lambda^\varepsilon \mathbf{C}[[\lambda]]/\lambda^{N+\varepsilon}\mathbf{C}[[\lambda]]$ :

$$(x_1, y_1, \cdots, x_k, y_k) \mapsto \rho = \frac{C_k}{D_k} \bmod \lambda^{-2k-1}, \quad \text{where}$$

$$\begin{pmatrix} 1 & 0 \\ \lambda x_1 & 1 \end{pmatrix} \begin{pmatrix} 1 & y_1 \\ 0 & 1 \end{pmatrix} \cdots \begin{pmatrix} 1 & 0 \\ \lambda x_k & 1 \end{pmatrix} \begin{pmatrix} 1 & y_k \\ 0 & 1 \end{pmatrix} = \begin{pmatrix} A_k & C_k \\ B_k & D_k \end{pmatrix},$$

$(x_1, y_1, \cdots x_{k+1}) \mapsto \rho' = \frac{A'_k}{B'_k} \bmod \lambda^{-2k-2}$, where

$$\begin{pmatrix} 1 & 0 \\ \lambda x_1 & 1 \end{pmatrix} \cdots \begin{pmatrix} 1 & 0 \\ \lambda x_{k+1} & 1 \end{pmatrix} = \begin{pmatrix} A'_k & C'_k \\ B'_k & D'_k \end{pmatrix}.$$

$(y_1, x_2, \cdots, x_{k+1}) \mapsto \sigma = \frac{\bar{B}_k}{\bar{A}_k} \bmod \lambda^{-2k}$, where

$$\begin{pmatrix} 1 & y_1 \\ 0 & 1 \end{pmatrix} \cdots \begin{pmatrix} 1 & 0 \\ \lambda x_{k+1} & 1 \end{pmatrix} = \begin{pmatrix} \bar{A}_k & \bar{C}_k \\ \bar{B}_k & \bar{D}_k \end{pmatrix}.$$

$(y_1, x_2, \cdots, x_k, y_k) \mapsto \sigma' = \dfrac{\bar{D}'_k}{\bar{C}'_k} \mod \lambda^{-2k+1}$, where

$$\begin{pmatrix} 1 & y_1 \\ 0 & 1 \end{pmatrix} \cdots \begin{pmatrix} 1 & y_k \\ 0 & 1 \end{pmatrix} = \begin{pmatrix} \bar{A}'_k & \bar{C}'_k \\ \bar{B}'_k & \bar{D}'_k \end{pmatrix}.$$

This defines $\widehat{b}_-$-module injections $\mathbf{C}[\sigma_i] = \mathbf{C}[\widehat{B}_-/\widehat{B}_- \cap \widehat{B}^w_-] \hookrightarrow \mathbf{C}[x_i, y_i]$ in the two first cases, and $\mathbf{C}[\rho_i] = \mathbf{C}[\widehat{B}_-/\widehat{B}_- \cap \widehat{B}^w_-] \hookrightarrow \mathbf{C}[x_i^{-1}, y_i^{-1}]$ in the two last ($\rho_i = \sum_{i \geq 1} \rho_i \lambda^{-i}, \sigma_i = \sum_{i \geq 0} \sigma_i \lambda^{-i}$).

## 4. Computation of cohomologies.

Let us consider a point $b_-(\widehat{B}_- \cap \widehat{B}^w_-)$ where the mapping $\widehat{B}_-/\widehat{B}_- \cap \widehat{B}^w_- \to (\mathbf{C}^*)^N$ is regular and the local ring at this point : it is isomorphic to the coinduced module $U\widehat{b}_- \otimes_{U(\widehat{b}_- \cap \widehat{b}^w_-)} \mathbf{C}$. We then have the sequence of maps

$$H^1(\widehat{b}_-, \mathbf{C}[\widehat{B}_-/\widehat{B}_- \cap \widehat{B}^w_-]) \to H^1(\widehat{b}_-, \mathbf{C}[x_i^{\pm 1}, y_i^{\pm 1}]) \to H^1(\widehat{b}_-, (U\widehat{b}_- \otimes_{U(\widehat{b}_- \cap \widehat{b}^w_-)} \mathbf{C})^*)$$

The last space is isomorphic to $H^1(\widehat{b}_- \cap \widehat{b}^w_-)$, by Shapiro's lemma. If we find representatives for classes in this space by functions in $\mathbf{C}[\widehat{B}_-/\widehat{B}_- \cap \widehat{B}^w_-]$ their images in $H^1(\widehat{b}_-, \mathbf{C}[x_i^{\pm 1}, y_i^{\pm 1}])$ will not vanish. Pose $s_w = \widehat{b}_-/\widehat{b}_- \cap \widehat{b}^w_-$. Then $H^1(s_w) = (s_w/[s_w, s_w])^*$. Then $s_w/[s_w, s_w]$ is spanned by the classes of the $\begin{pmatrix} \lambda^{-i} & 0 \\ 0 & -\lambda^{-i} \end{pmatrix}, 0 \leq i \leq 2k, 2k+1, 2k, 2k-1$ in the four cases we considered, and $H^1(s_w)$ is spanned by the forms $\varphi_i : x(\lambda) \mapsto \frac{1}{2} \text{res}_\infty \frac{d\lambda}{\lambda} \text{tr} \begin{pmatrix} \lambda^i & 0 \\ 0 & -\lambda^i \end{pmatrix} x(\lambda)$.

Let us show now how an element of $H^1(s_w)$ can give a cohomology class in $H^1(\widehat{b}_-, \text{Fun}(\widehat{B}_-/S_w))$. Let us choose a section $\sigma$ of the projection $\widehat{B}_- \to \widehat{B}_-/S_w$. For $X \in \widehat{b}_-$ and $x \in \widehat{B}_-/S_w$ we write $(1+\varepsilon X)\sigma(x) = \sigma(x_\varepsilon)(1+\varepsilon s(X, x)+o(\varepsilon))$, with $S$ linear in $X$ with values in $s_w$. Then for $\varphi \in (s_w/[s_w, s_w])^*$ we pose $f_X(x) = \langle \varphi, s(X, x) \rangle$ ; it is a 1-cocycle of $\widehat{b}_-$ in $\text{Fun}(\widehat{B}_-/S_w)$. Let us choose for $\sigma$ the maps $\rho, \rho' \mapsto \begin{pmatrix} 1 & \tilde{\rho} \\ 0 & 1 \end{pmatrix}, \begin{pmatrix} 1 & \tilde{\rho}' \\ 0 & 1 \end{pmatrix}$ and $\sigma, \sigma' \mapsto \begin{pmatrix} 1 & 0 \\ \tilde{\sigma} & 1 \end{pmatrix}, \begin{pmatrix} 1 & 0 \\ \tilde{\sigma}' & 1 \end{pmatrix}$. Then the cocycles corresponding to $\varphi_i$ are the maps $f_X^i(\rho) = \frac{1}{2}\text{res}_\infty \frac{d\lambda}{\lambda} \text{tr} \begin{pmatrix} \lambda^i & 0 \\ 0 & -\lambda^i \end{pmatrix} \begin{pmatrix} 1 & -\tilde{\rho} \\ 0 & 1 \end{pmatrix} X \begin{pmatrix} 1 & \tilde{\rho} \\ 0 & 1 \end{pmatrix}$, same formula for $\rho'$, $f_X^i(\sigma) = \frac{1}{2}\text{res}_\infty \frac{d\lambda}{\lambda} \text{tr} \begin{pmatrix} \lambda^i & 0 \\ 0 & -\lambda^i \end{pmatrix} \begin{pmatrix} 1 & 0 \\ -\tilde{\sigma} & 1 \end{pmatrix} X \begin{pmatrix} 1 & 0 \\ \tilde{\sigma} & 1 \end{pmatrix}$.

More explicitly, we have the cocycles $f_X^i$ of $\widehat{b}_-$ in $\mathbf{C}[\rho_1, \cdots, \rho_n]$, such that $f_{h\lambda^{-i}}^i = 1$, $f_{f\lambda^{-j}}^i = -\rho_{i-j}$, $f_{h\lambda^{-k}}^i = f_{e\lambda^{-k'}}^i = 0$ if $k \neq i$, $k' \geq 1$, and the cocycles $g_X^i$ of $\widehat{b}_-$ in $\mathbf{C}[\sigma_0, \cdots, \sigma_n]$, such that $g_{h\lambda^{-i}}^i = 1$, $g_{e\lambda^{-j}}^i = \sigma_{i-j}$, $g_{h\lambda^{-k}}^i = g_{f\lambda^{-k'}}^i = 0$ if $k \neq i$, $k' \geq 0$.

**Remark.** The action of $\widehat{b}_-$ and the cocycles can be expressed as follows, on the manifold $\widehat{B}_-w\widehat{B}_- \cap \widehat{N}_+$ : the action of $X \in \widehat{b}_-$ at the point $g \in \widehat{B}_-w\widehat{B}_- \cap \widehat{N}_+$ is $\mathcal{L}_X g = g(g^{-1}Xg)_+$ ; to express the cocycle corresponding to $\lambda^i h(i \geq 0)$ we use the embedding $\widehat{N}_+ \hookrightarrow \widehat{G}/\widehat{B}_-$ ; the formula is then $f_X^i(g) = \mathrm{res}_\infty \frac{d\lambda}{\lambda} \mathrm{tr}(\lambda^i h g^{-1} X g), X \in \widehat{b}_-$. This gives formulas for the cocycles in terms of $x_i$ and $y_i$ (and not their inverses).

## 5. Computation of $T$ and $I$ in cohomologies.

Let $\psi^\pm(x_1, y_1, \cdots, x_k, y_k)$ belong to $Z^1(\widehat{b}_-, \mathbf{C}[x_1, \cdots, y_k])$ ; then we have cocycles $T\psi$ and $I\psi$ in $Z^1(\widehat{b}_-, \mathbf{C}[x_1, \cdots, y_{k+1}])$, defined by $(I\psi)^\pm(x_1, \cdots, y_{k+1}) = \psi^\pm(x_1, \cdots, y_k)$, and $(T\psi)^\pm(x_1, \cdots, y_{k+1}) = \psi^\pm(x_2, \cdots, y_{k+1})$. $I\psi$ is the image of $\psi$ by the $\widehat{b}_-$-module map $I : \mathbf{C}[x_1, \cdots, y_k] \to \mathbf{C}[x_1, \cdots, y_{k+1}]$, $x_i, y_i \mapsto x_i, y_i$. Recall now the maps $\mathbf{C}[\rho_1, \cdots, \rho_{2k}] \to \mathbf{C}[x_1, \cdots, y_k]$, $\mathbf{C}[\rho_1, \cdots, \rho_{2k+2}] \to \mathbf{C}[x_1, \cdots, y_{k+1}]$. Then $I$ corresponds to the map $\mathbf{C}[\rho_1, \cdots, \rho_{2k}] \to \mathbf{C}[\rho_1, \cdots, \rho_{2k+2}], \rho_i \mapsto \rho_i$. Indeed, recalling the notations of section 3, we have

$$\begin{vmatrix} C_k & C_{k+1} \\ D_k & D_{k+1} \end{vmatrix} = \begin{vmatrix} C_k & y_{k+1}A_k + (1 + x_{k+1}y_{k+1})C_k \\ D_k & y_{k+1}B_k + (1 + x_{k+1}y_{k+1})D_k \end{vmatrix} = -y_{k+1} .$$

Then $\frac{C_k}{D_k} - \frac{C_{k+1}}{D_{k+1}} = \frac{-y_{k+1}}{D_k D_{k+1}} \in \lambda^{-2k-1}\mathbf{C}[[\lambda^{-1}]]$, since $D_k$ and $D_{k+1}$ have respective degrees in $\lambda$, $k$ and $k+1$. It means that the classes in $\lambda^{-1}\mathbf{C}[[\lambda^{-1}]]/\lambda^{-2k-1}\mathbf{C}[[\lambda^{-1}]]$ of $\rho_k$ and $\rho_{k+1}$ are the same. We deduce from this that the map $I$ corresponds to the map $\mathbf{C}[\widehat{B}_-/\widehat{B} \cap \widehat{B}^{w_k}] \to \mathbf{C}[\widehat{B}_-/\widehat{B} \cap \widehat{B}^{w_{k+1}}]$, $w_k = \begin{pmatrix} \lambda^{-k} & 0 \\ 0 & -\lambda^k \end{pmatrix}$, induced by the natural projection, and that it induces in cohomology the map $I : H^1(s_{w_k}) \to H^1(s_{w_{k+1}})$ coming from the injection $s_{w_{k+1}} \to s_{w_k}$. In particular it maps the class of $f^i \in Z^1(\widehat{b}_-, \mathbf{C}[\rho_1, \cdots, \rho_k])$ to the class of $f^i \in Z^1(\widehat{b}_-, \mathbf{C}[\rho_1, \cdots, \rho_{k+1}])$.

Let us now compute $T$ in cohomology. Let us consider the map $T : \mathbf{C}[x_1, \cdots, y_k] \to \mathbf{C}[x_1, \cdots, y_{k+1}]$, $x_i, y_i \mapsto x_{i+1}, y_{i+1}$. It is the composition of the maps $T_1 : \mathbf{C}[x_1, \cdots, y_k] \to \mathbf{C}[y_1, x_2, \cdots, y_{k+1}]$, $x_i, y_i \mapsto x_{i+1}, y_{i+1}$, and $T_2 : \mathbf{C}[y_1, x_2, \cdots y_{k+1}] \to \mathbf{C}[x_1, \cdots, y_{k+1}]$, $x_i$, $y_i \mapsto x_i, y_i$ .

Let us define algebra isomorphisms $T_1'$ from $\mathbf{C}[\sigma_0, \cdots, \sigma_{2k}]$ to $\mathbf{C}[\rho_1, \cdots, \rho_{2k}] \otimes \mathbf{C}[y_1^{-1}]$ by $\sigma = \sum_{i=0}^{2k} \sigma_i \lambda^{-i} = \frac{1}{y_1 + \rho}$, where $\rho = \sum_{i=1}^{2k} \rho_i \lambda^{-i}$, and $T_2'$ from $\mathbf{C}[\rho_1', \cdots, \rho_{2k+2}']$ to $\mathbf{C}[\sigma_0, \cdots, \sigma_{2k}] \otimes \mathbf{C}[x_1^{-1}]$ by $\rho' = \frac{1}{\lambda x_1 + \sigma}$, where $\rho' = \sum_{i=1}^{2k+2} \rho_i' \lambda^{-i}$. The inverses of these correspond to $T_1$ and $T_2$.

$T_1'$ and $T_2'$ are $\widehat{n}_-$-module morphisms if on $\mathbf{C}[\rho_1, \cdots, \rho_{2k}] \otimes \mathbf{C}[y_1^{-1}]$, $e\lambda^{-n-1}, h\lambda^{-n}$, $f\lambda^{-n}$ act respectively as the vector fields

$$\frac{\partial}{\partial \rho_{n+1}}, 2y_1 \frac{\partial}{\partial \rho_n} + 2\sum_{k \geq 1} \rho_k \frac{\partial}{\partial \rho_{n+k}}, -\sum_{k \geq 0}(\sum_{i=0}^k \rho_i \rho_{k-i}) \frac{\partial}{\partial \rho_{n+k}}$$

(in the last expression we set $\rho_0 = y_1$), and on $\mathbf{C}[\sigma_0, \cdots, \sigma_{2k}] \otimes \mathbf{C}[x_1^{-1}]$, by

$$-\sum_{k \geq -2}(\sum_{i \geq -1} \sigma_i \sigma_{k-i}) \frac{\partial}{\partial \sigma_{k+n+1}}, -2\sum_{k \geq -1} \sigma_k \frac{\partial}{\partial \sigma_{n+k}}, \frac{\partial}{\partial \sigma_n} ,$$

where in the two first expressions we set $\sigma_{-1} = x_1$.

Then we have maps of cohomologies $H^1(\widehat{b}_-, \mathbf{C}[\rho_1, \cdots, \rho_{2k}]) \to H^1(\widehat{b}_-, \mathbf{C}[\rho_1, \cdots, \rho_{2k}] \otimes \mathbf{C}[y_1^{-1}]) \xrightarrow{T_1'^{-1}} H^1(\widehat{b}_-, \mathbf{C}[\sigma_0, \cdots, \sigma_{2k}])$ and $H^1(\widehat{b}_-, \mathbf{C}[\sigma_0, \cdots, \sigma_{2k}]) \to H^1(\widehat{b}_-, \mathbf{C}[\sigma_0, \cdots, \sigma_{2k}] \otimes \mathbf{C}[x_1^{-1}]) \xrightarrow{T_2'^{-1}} H^1(\widehat{b}_-, \mathbf{C}[\rho_1', \cdots, \rho_{2k+2}'])$, where the two first maps associate to the cocycle $\varphi$ in $\mathrm{Hom}(\widehat{b}_-, \mathbf{C}[\rho_1, \cdots, \rho_{2k}])$ or $\mathrm{Hom}(\widehat{b}_-, \mathbf{C}(\sigma_0, \cdots, \sigma_{2k}])$, the cocycle $\bar{\varphi}$ defined by $\bar{\varphi}(e\lambda^{-1}) = \varphi(e\lambda^{-1}) \otimes 1, \bar{\varphi}(f) = \varphi(f) \otimes 1$.

The image by $^-$ of $f^i \in Z^1(\widehat{b}_-, \mathbf{C}[\rho_1, \cdots, \rho_{2k}])$ is the cocycle $\bar{f}^i$, such that $\bar{f}^i_{h\lambda^{-i}} = 1$, $\bar{f}^i_{f\lambda^{-j}} = -\rho_{i-j}$, $\bar{f}^i_{f\lambda^{-i}} = -y_1, \bar{f}^i_{f\lambda^{-k''}} = \bar{f}^i_{h\lambda^{-k'''}} = \bar{f}^i_{e\lambda^{-k'}} = 0$ if $k''' \neq i$, $k' \geq 0$, $k'' > i$ ; the image by $^-$ of $g^i \in Z^1(\widehat{b}_-, \mathbf{C}[\sigma_0, \cdots, \sigma_{2k}])$ is the cocycle $\bar{g}^i$, such that

$$\bar{g}^i_{h\lambda^{-i}} = 1, \bar{g}^i_{e\lambda^{-j}} = \sigma_{i-j}, \bar{g}^i_{e\lambda^{-i-1}} = x_1, \bar{g}^i_{h\lambda^{-k'}} = \bar{g}^i_{e\lambda^{-k''}} = \bar{g}^i_{f\lambda^{-k'''}} = 0$$

if $k' \neq i$, $k'' > i+1$, $k''' \geq 0$.

Let us compute the classes of $T_1'^{-1}(f^i)$ and $T_2'^{-1}(g^i)$. The mapping $H^1(\widehat{b}_-, \mathrm{Fun}(\widehat{B}_-/S^w)) \to H^1(s^w)$ is defined as follows ([G]) : to the cocycle $(X \mapsto \varphi_X) \in Z^1(\widehat{b}_-, \mathrm{Fun}(\widehat{B}_-/S^w))$, we associate the form $X \in s^w \mapsto \varphi_X(eS^w)$. This operation is impossible here because $eS^w$ corresponds to $\sigma_0 = \cdots = \sigma_{2k} = 0$ in the first case, and $\rho_1' = \cdots = \rho_{2k+2}' = 0$ in the second, where the function $y_1 = \sigma_0^{-1}$ (resp. $x_1 = \rho_1'^{-1}$) is not defined. So we will use evaluation at another point of $\widehat{B}_-/S^w$. Let $b_- \in \widehat{B}_-$ ; the stabilizer of $b_-S^w$ is $b_-S^w b_-^{-1}$, and the natural map $H^1(\widehat{b}_-, \mathrm{Fun}(\widehat{B}_-/S^w)) \to H^1(b_-S^w b_-^{-1})$ is $(X \mapsto \varphi_X) \mapsto (X \in b_-s^w b_-^{-1} \mapsto \varphi_X(b_-S^w))$ ; finally we have the natural map

$$H^1(\widehat{b}_-, \mathrm{Fun}(\widehat{B}_-/S^w)) \to H^1(s^w), (X \mapsto \varphi_X) \mapsto (X \in s^w \mapsto \varphi_{b_-Xb_-^{-1}}(b_-S^w)) .$$

For $T_1'^{-1}(f^i)$, we choose $b_- = \begin{pmatrix} 1 & 0 \\ \alpha & 1 \end{pmatrix}, \alpha \neq 0$. Then

$$\varphi_{b_-(\lambda^{-n}f)b_-^{-1}} = -\rho_{i-n}, \varphi_{b_-(\lambda^{-n}h)b_-^{-1}} = \delta_{in} - 2\alpha\rho_{i-n} , \varphi_{b_-(\lambda^{-n-1}e)b_-^{-1}} =$$

$$= -\alpha\delta_{i,n+1} - \alpha^2\rho_{i-n-1}$$ (posing again $\rho_0 = y_1^{-1} = \sigma_1^{-1}$). We are only interested in $\varphi_{b_-(\lambda^{-n}h)b_-^{-1}}(b_-S^w)$ and it is $-\delta_{in}$. So $T_1$ maps the class of $f^i$ to the class of $-g^i$.

For $T_2'^{-1}(g^i)$, we choose $b_- = \begin{pmatrix} 1 & \beta \\ 0 & 1 \end{pmatrix}, \beta \neq 0$. Then $\varphi_{b_-(\lambda^{-n}h)b_-^{-1}} =$

$$= \delta_{ni} - 2\beta\sigma_{i-n}, \varphi_{b_-(\lambda^{-n}e)b_-^{-1}} = \sigma_{i-n}, \varphi_{b_-(\lambda^{-n}f)b_-^{-1}} = \beta\delta_{in} - \beta^2\sigma_{i-n}.$$

So $T_2$ maps the class of $g^i$ to the class of $-f^i$.

In conclusion we see that in cohomology, $T - I$ is equal to zero. It means that for each $k$, we have $k$ integrals of motion corresponding to polynomials in $\mathbf{C}[x_1^{-1}, y_1^{-1}, \cdots, x_k^{-1}, y_k^{-1}]$.

We can repeat the reasoning for the other algebras $\mathbf{C}[y_1^{-1},\cdots,y_k^{-1}]$, etc., and obtain for cohomologies of the diagram of algebras

$$
\begin{array}{ccccccc}
\mathbf{C}[x_1^{-1}] & \xrightarrow{I} & \cdots & \mathbf{C}[x_1^{-1},\cdots,y_k^{-1}] & \xrightarrow{I} & \mathbf{C}[x_1^{-1},\cdots,x_{k+1}^{-1}] & \longrightarrow \\
& \nwarrow \nearrow^T & & \nwarrow \nearrow^T & & & \\
\mathbf{C} & T \times & \cdots & T \times & & \times & \cdots \\
& \swarrow \searrow & & \swarrow \searrow & & & \\
\mathbf{C}[y_1^{-1}] & \xrightarrow[I]{} & \cdots & \mathbf{C}[y_2^{-1},\cdots,x_{k+1}^{-1}] & \xrightarrow{I} & \mathbf{C}[y_2^{-1},\cdots,y_{k+1}^{-1}] & \longrightarrow
\end{array}
$$

the diagram

$$
\begin{array}{ccccccc}
\mathbf{C}h & \xrightarrow{1} & \cdots & \bigoplus_{i=0}^{2k} \mathbf{C}\lambda^i h & \xrightarrow{1} & \bigoplus_{i=0}^{2k+1} \mathbf{C}\lambda^i h & \\
& \nearrow^1 & & \nearrow^{-1} & & & \\
\mathbf{C}h & \times_{-1} & \cdots & \times_{-1} & & \cdots & \quad (**) \\
& \searrow_1 & & & & & \\
\mathbf{C}h & \xrightarrow[1]{} & \cdots & \bigoplus_{i=0}^{2k} \mathbf{C}\lambda^i h & \xrightarrow[1]{} & \bigoplus_{i=0}^{2k+1} \mathbf{C}\lambda^i h &
\end{array}
$$

Let us denote the space of integrals of motion in $x_1^{-1},\cdots,y_k^{-1}$ by $IM(x_1^{-1},\cdots,y_k^{-1})$: it is the subspace of $\mathbf{C}[x_1^{-1},\cdots,y_k^{-1}]/\mathrm{Im}(T-1)$ of classes of polynomials $P$ whose Poisson brackets with $\sum_{i=1}^k x_i, \sum_{i=1}^k y_i$ are in $\mathrm{Im}(T-1)$. Let us remark also that the part $\mathbf{C}h$ of the cohomology corresponds to a trivial cocycle $\psi^\pm = 0, \psi^0 = 1$. We summarize our results:

**Proposition.—** *The spaces $IM$ $(x_1^{-1},\cdots,y_{k+1}^{-1})$, $IM$ $(x_1^{-1},\cdots,x_{k+2}^{-1})$, $IM(y_2^{-1},\cdots,x_{k+2}^{-1})$, $IM(y_2^{-1},\cdots,y_{k+2}^{-1})$ are graded linear spaces of respective dimensions $2(k-1), 2k-1, 2(k-1), 2k-1$. Setting $IM(x_1^{-1},\cdots,y_k^{-1}) = \bigoplus_{s=1}^{2k-2} \mathbf{C}I_s(x_1^{-1},\cdots,y_k^{-1})$, etc., the identity and the translation induce maps between the spaces $IM(x_1^{-1},\cdots,y_k^{-1})$, $IM(x_1^{-1},\cdots,x_{k+1}^{-1})$, etc., analogous to $(**)$.*

## 6. Expression of the integrals of motion.

Let us consider the cocycle $\lambda^{2k+1}h \in H^1(\widehat{b}_-, \mathbf{C}[x_1^{-1}, \cdots, x_{k+1}^{-1}])$. By the map $\mathbf{C}[x_1^{-1}, \cdots, x_{k+1}^{-1}] \simeq \mathbf{C}[\rho'_1, \cdots, \rho'_{2k+1}]$ (sect. 3), $\lambda^{2k+1}h$ is identified with the class of $f^{2k+1}$. Consider now the map $T_1 : \mathbf{C}[x_1^{-1}, \cdots, x_{k+1}^{-1}] \to \mathbf{C}[y_1^{-1}, \cdots, x_{k+2}^{-1}], x_i, y_i \mapsto x_{i+1}, y_{i+1}$. The class of the cocycle $T_1 f^{2k+1} \in Z^1(\widehat{b}_-, \mathbf{C}[y_1^{-1}, \cdots, x_{k+2}^{-1}])$, defined by $(T_1 f^{2k+1})(Q_\pm) = T_1(f^{2k+1}(Q_\pm))$ coincides with the class of $-g^{2k+1}$.

Let us write $T_1 f^{2k+1} + g^{2k+1} = d\varphi$. By the identification of $\mathbf{C}[y_1^{-1}, \cdots, x_{k+2}^{-1}]$ with $\mathbf{C}[\sigma'_0, \cdots, \sigma'_{2k+1}]$, we see that we should have $\mathcal{L}_{h\lambda^{-j}}\varphi = 2\delta_{j,2k+1}$, $\mathcal{L}_{e\lambda^{-i}}\varphi = \sigma_{2k+1-i}$, $\mathcal{L}_{f\lambda^{-i}}\varphi = -(\frac{1}{\sigma})_{2k+1-i}$. The third equality gives us equations satisfied by
$$\varphi\left(\begin{pmatrix} 1 & 0 \\ \sigma_0 + \sigma_1\lambda^{-1} & \cdots & 1 \end{pmatrix}\right) = \bar{\varphi}(\sigma_0, \sigma_1, \cdots) : \frac{\partial\bar{\varphi}}{\partial\sigma_i} = -(\frac{1}{\sigma})_{2k+1-i}.$$
Write now $\sigma_0 + \sigma_1\lambda^{-1} + \cdots = \exp(\ln\sigma_0 + \sum_{i\geq 1}\xi_i\lambda^{-i})$. We have $\sum_{\alpha\geq 0}\frac{\partial\sigma_i}{\partial\xi_\alpha}\frac{\partial\xi_\alpha}{\partial\sigma_k} = \delta_{ik}$ (with $\xi_0 = \ln\sigma_0$), so $\sum_{\alpha\geq 0}\sigma_{i-\alpha}\frac{\partial\xi_\alpha}{\partial\gamma_k} = \delta_{ik}$ and $\frac{\partial\xi_\alpha}{\partial\sigma_k} = (\frac{1}{\sigma})_{\alpha-k}$. We thus obtain $\bar{\varphi} = -\xi_{2k+1}$ and check that the two first equations on $\bar{\varphi}$ are satisfied.

Consider now the map $T_2 : \mathbf{C}[y_1^{-1}, \cdots, x_{k+2}^{-1}] \to \mathbf{C}[x_1^{-1}, \cdots, x_{k+2}^{-1}], x_i, y_i \mapsto x_i, y_i$. The class of the cocycle $T_2 g^{2k+1} \in Z^1(\widehat{b}_-, \mathbf{C}[x_1^{-1}, \cdots, x_{k+2}^{-1}])$ defined by $(T_2 g^{2k+1})(Q_\pm) = T_2(g^{2k+1}(Q_\pm))$ coincides with the class of $-f^{2k+1}$. Let us write $T_2 g^{2k+1} + f^{2k+1} = d\psi$. After the identification of $\mathbf{C}[x_1^{-1}, \cdots, x_{k+2}^{-1}]$ with $\mathbf{C}[\rho_1, \cdots, \rho_{2k+3}]$, we obtain the conditions on $\psi$, $\mathcal{L}_{h\lambda^{-j}}\psi = 2\delta_{j,2k+1}$, $\mathcal{L}_{e\lambda^{-i}}\psi = (\frac{1}{\rho})_{2k+1-i}$, $\mathcal{L}_{f\lambda^{-i}}\psi = -\rho_{2k+1-i}$. Posing $\bar{\psi}(\rho_1, \rho_2, \cdots) = \psi\left(\begin{pmatrix} 1 & \rho_1\lambda^{-1} + \cdots \\ 0 & 1 \end{pmatrix}\right)$, the second condition is translated into $\frac{\partial\bar{\psi}}{\partial\rho_i} = (\frac{1}{\rho})_{2k+1-i}$.

Let us write $\rho_1\lambda^{-1} + \rho_2\lambda^{-2} + \cdots = \exp(\ln\lambda^{-1} + \ln\rho_1 + \sum_{\alpha\geq 1}\eta_\alpha\lambda^{-\alpha})$, then $\sum_{\alpha\geq 0}\frac{\partial\rho_i}{\partial\eta_\alpha}\frac{\partial\eta_\alpha}{\partial\rho_k} = \delta_{ik}$ (with $\eta_0 = \ln\rho_1$), so $\sum_{\alpha\geq 0}\rho_{i-\alpha}\frac{\partial\eta_\alpha}{\partial\rho_k} = \delta_{ik}$ and $\frac{\partial\eta_\alpha}{\partial\rho_k} = (\frac{1}{\rho})_{\alpha-k}$. We thus obtain $\bar{\psi} = \eta_{2k+1}$, and check that the other conditions on $\bar{\psi}$ are satisfied.

We have finally $T_2 T_1 f^{2k+1} + T_2 g^{2k+1} = d(-T_2\xi_{2k+1})$, and $T_2 g^{2k+1} + f^{2k+1} = d(\eta_{2k+1})$, so that $Tf^{2k+1} - f^{2k+1} = d(-T_2\xi_{2k+1} - \eta_{2k+1})$. $T_2\xi_{2k+1} + \eta_{2k+1}$ is the conserved density corresponding to cocycle $\lambda^{2k+1}h$, it is expressed in terms of $x_i$ and $y_i$ as the coefficient of $\lambda^{-2k-1}$ in the expansion of $\ln\frac{\bar{B}_{k+1}}{B'_{k+1}}$. Similarly, we can show that the conserved densities representing the class $h\lambda^{2k+1}$ are proportional to

$$\operatorname{res}_{\lambda=\infty} \frac{d\lambda}{\lambda}\lambda^{2k+1}\ln\frac{\bar{B}_{k+1}}{B'_{k+1}}, \text{ or } \operatorname{res}_{\lambda=\infty} \frac{d\lambda}{\lambda}\lambda^{2k+1}\ln\frac{C_{k+1}}{C'_{k+1}},$$

and the conserved densities representing the class $h\lambda^{2k}$ are proportional to

$$\operatorname{res}_{\lambda=\infty} \frac{d\lambda}{\lambda}\lambda^{2k}\ln\frac{A'_k}{\bar{A}_{k+1}}, \text{ or } \operatorname{res}_{\lambda=\infty} \frac{d\lambda}{\lambda}\lambda^{2k}\ln\frac{\bar{D}'_{k+1}}{D_{k+1}},$$

in the notations of sect. 3.

**Theorem.**— *The conserved density of classical lattice sine-Gordon representing the class of $h\lambda^k$ is proportional to the coefficient of $\lambda^{-k}$ in the expansion of*

$$\ln(\lambda x_1/\lambda x_1 + 1/y_1 + \cdots + 1/\lambda x_N) + \ln(y_1/y_1 + 1/x_2 + \cdots + 1/y_N)$$

*with $N \geq 2(n+1)$ (we use notations of continued fractions); this coefficient becomes independent of $N$ in this region.*

So these classes can also be represented increasing the indices of $\bar{B}_{k+1}, \cdots$, e.g. $h\lambda^{2k+1}$ is also represented by $\text{res}_\infty \frac{d\lambda}{\lambda} \lambda^{2k+1} \ln \frac{\bar{B}_{k'+1}}{B'_{k'+1}}, k' \geq k$, etc. The first integrals of motion are $\sum_i (x_i y_i)^{-1} + (y_i x_{i+1})^{-1}$ and $\sum_i (x_i y_i^2 x_{i+1})^{-1} + \frac{1}{2}(x_i^2 y_i^2)^{-1} + \sum_i (y_i x_{i+1}^2 y_{i+1})^{-1} + \frac{1}{2}(y_i^2 x_{i+1}^2)^{-1}$.

**Remark.** Assume that these integrals of motions (here denoted by $I_k^{c\ell}$) admit quantizations $I_k$. Consider the case where $q$ is as root of 1, $q^n = 1$. Then it is natural (assuming commutativity of $I_k$'s) to compute $\text{tr}(\prod_{k \geq 0} I_k^{\alpha_k})$ ; the trace is taken in the module $\mathbf{C}[x_i, y_i]/(x_i^n = X_i, y_i^n = Y_i)$. This trace is a function of $X_i, Y_i$, and (recalling the Poisson action of an algebra on its center at roots of 1) we will have,

$$\{\sum x_i, \text{tr}(\prod_{k \geq 0} I_k^{\alpha_k})\} = \{\sum y_i, \text{tr}(\prod_{k \geq 0} I_k^{\alpha_k})\} = 0$$

(since the corresponding commutators are always zero). We deduce that the Poisson brackets of $(\sum x_i)^n$ and $(\sum y_i)^n$ with $\text{tr}(\prod_{k \geq 0} I_k^{\alpha_k})$ are also zero ; remark that the first quantities are $\sum X_i$ and $\sum Y_i$. Note that $\text{tr}(\prod_{k \geq 0} I_k^{\alpha_k})$ has the form $\sum_{s=1}^r \sum_{\alpha=1}^{k_s} \sum_{i_1 < \cdots < i_s} m_{1,s}^\alpha(i_1) \cdots m_{s,s}^\alpha(i_s)$, the $m_{i,k}^\alpha$ being polynomials in $X_i, Y_i$ and $P(i)$ denoting the polynomial $P$, translated by $i$, with the tensors $\sum_{\alpha=1}^{k_s} m_{1,s}^\alpha \otimes \cdots \otimes m_{s,s}^\alpha$ totally symmetric. The commutation of $\text{tr}(\prod_{k \geq 0} I_k^{\alpha_k})$ with $\sum_i X_i$ and $\sum_i Y_i$ imposes that the $m_{i,s}^\alpha$ are densities of the classical integrals of motion $I_k^{c\ell}(X_i, Y_i)$; substracting then to $\text{tr}(\prod_{k \geq 0} I_k^{\alpha_k})$ an appropriate polynomial in $I_k^{c\ell}(X_i, Y_i)$, we obtain an expression of the form $\sum_{s=1}^{r-1} \sum_{\alpha=1}^{k'_s} \sum_{i_1 < \cdots < i_s} n_{1,s}^\alpha(i_1) \cdots n_{s,s}^\alpha(i_s)$, commuting to $\sum_i X_i$ and $\sum_i Y_i$. By induction we see that $\text{tr}(\prod_{k \geq 0} I_k^{\alpha_k})$ is a polynomial in the classical integrals of motion $I_k^{c\ell}(X_i, Y_i)$.

## 7. An other lattice version of the sine-Gordon system.

An other way to discretise the sine-Gordon system is the following : we consider a system of variables $x_i^{\pm 1} (i \in \mathbf{Z})$ on the line, with $x_i x_j = q x_j x_i, i < j$, and define the screening actions to be $\bar{Q}_{\pm} = [\sum_i x_i^{\pm 1}, \cdot]_q$.

Classical limit of these operators are the vector fields $Q_{\pm}(1) = \pm \sum e^{\pm \xi_i}(\frac{\partial}{\partial \xi_i} + 2\sum_{j>i} \frac{\partial}{\partial \xi_j})$ (we pose $x_i = e^{\xi_i}$). Pose $H(0) = 2\sum \frac{\partial}{\partial \xi_i}$, and $[Q_+(2i+1), Q_-(1)] = H(2i+2), [H(2i), Q_{\pm}(1)] = \pm 2 Q_{\pm}(2i+1)$. This is an action of $\widehat{b}_-$ on $\mathbf{C}[x_1^{\pm 1}, \cdots, x_k^{\pm 1}]$.

Denote still by $Q_{\pm}(2i+1)$, $H(2i)$ the action of the above operators on $\mathbf{C}[x_1^{\pm 1}, \cdots, x_k^{\pm 1}]$ and by $\widehat{Q}_{\pm}(2i+1), \widehat{H}(2i)$ their action on $\mathbf{C}[x_0^{\pm 1}, \cdots, x_k^{\pm 1}]$. Then we have the formulas :

$$\widehat{Q}_+(2n+1) = e^{\tau_0}\frac{\partial}{\partial \tau_0} + e^{\tau_0}(H(0) + \cdots + H(2n)) + (Q_+(1) + \cdots + Q_+(2n+1))$$
$$- e^{2\tau_0}(Q_-(1) + \cdots + Q_-(2n-1)),$$

$$\widehat{Q}_-(2n+1) = -e^{-\tau_0}\frac{\partial}{\partial \tau_0} - e^{-\tau_0}(H(0) + \cdots + H(2n))$$
$$- e^{-2\tau_0}(Q_+(1) + \cdots + Q_+(2n-1)) + (Q_-(1) + \cdots + Q_-(2n+1))$$

$$\widehat{H}(2n) = 2\frac{\partial}{\partial \tau_0} + (2H(0) + \cdots + 2H(2n-2) + H(2n)),$$
$$- 2e^{\tau_0}(Q_-(1) + \cdots + Q_-(2n-1)) + 2e^{-\tau_0}(Q_+(1) + \cdots + Q_+(2n+1)).$$

This allows to show that the values at the point $x_1 = \cdots = x_N = 1$ of these vector fields are given by

$$H(\lambda) = \sum_{i=0}^{\infty} H(2i)_{|x_i=1} \lambda^{2i} = \sum_{k=1}^{N} \frac{(1+\lambda)^{2k-1} + (1-\lambda)^{2k-1}}{(1-\lambda^2)^k} \frac{\partial}{\partial \tau_k},$$

$$Q^+(\lambda) = \sum_{i=0}^{\infty} Q^+(2i+1)_{|x_i=1} \lambda^{2i+1} = \sum_{k=1}^{N} \frac{(1+\lambda)^{2k-1} - (1-\lambda)^{2k-1}}{(1-\lambda^2)^k} \frac{\partial}{\partial \tau_k},$$

$$Q^-(\lambda) = \sum_{i=0}^{\infty} Q^-(2i+1)_{|x_i=1} \lambda^{2i+1} = -Q^+(\lambda).$$

These vector fields span the tangent space at $x_1 = \cdots = x_N = 1$; the stabilizer of their action is the subalgebra of $\widehat{b}_+$

$$a_N = a + (1-\lambda)^N \widehat{b}_+ + s,$$

where $a = \underset{i \geq 0}{\oplus} \mathbf{C}\lambda^i(e + \lambda f)$ is the principal commutative subalgebra, and

$$s = \underset{i \geq 0}{\oplus} \mathbf{C}[t^i(1+t)^N(ft + h - \frac{e}{t}) + (-t)^i(1-t)^N(-ft + h + \frac{e}{t})] \text{ (here } t^2 = \lambda).$$

We have $[a, s] \subset s, [s, s] \subset (1-\lambda)^N \widehat{n}_+ \cap a$, and we can see that $a_N/[a_N, a_N]$ is spanned by the classes of $\lambda^i(e + \lambda f), i = 0, \cdots, N-1$.

We have a map $H^1(\widehat{b}_+, \mathbf{C}[x_1^{\pm 1}, \cdots, x_N^{\pm 1}]) \to H^1(a_N)$, since by Shapiro's lemma, we have an isomorphism $H^1(\widehat{b}_+, \mathbf{C}[[x_1 - 1, \cdots, x_N - 1]]) \simeq H^1(a_N)$. The injection $\mathbf{C}[x_1^{\pm 1}, \cdots, x_N^{\pm 1}] \hookrightarrow \mathbf{C}[x_1^{\pm 1}, \cdots, x_{N+1}^{\pm 1}], x_i \mapsto x_i$, is a morphism of $\widehat{b}_+$-modules and induces the natural map $H^1(a_N) \to H^1(a_{N+1})$ in cohomology.

Let us now compute the injection $T : \mathbf{C}[x_1^{\pm 1}, \cdots, x_N^{\pm 1}] \to \mathbf{C}[x_0^{\pm 1}, \cdots, x_N^{\pm 1}], x_i \mapsto x_i$ in cohomology. The image of the cocycle $\psi$ will be defined by $(T\psi)(Q_\pm) = T(\psi(Q_\pm))$. By the above formulas for $\widehat{Q}_\pm, \widehat{H}$, we then have

$$(T\psi)(Q_+(2n+1)) = x_0[\psi(H(0)) + \cdots + \psi(H(2n))] + [\psi(Q_+(1)) + \cdots$$
$$+ \psi(Q_+(2n+1)] - [\psi(Q_-(1)) + \cdots + \psi(Q_-(2n-1))]$$
$$(T\psi)(Q_-(2n+1)) = -x_0^{-1}[\psi(H(0)) + \cdots + \psi(H(2n))] + [\psi(Q_-(1)) + \cdots$$
$$+ \psi(Q_-(2n+1))] - [\psi(Q_+(1)) + \cdots + \psi(Q_+(2n-1))]$$
$$(T\psi)(H(2n)) = [2\psi(H(0)) + \cdots + 2\psi(H(2n-1)) + \psi(H(2n))] - [\psi(Q_-(1)) + \cdots$$
$$+ \psi(Q_-(2n-1))] + 2[\psi(Q_+(1)) + \cdots + \psi(Q_+(2n+1))] \; .$$

We then compute

$$(T\psi)(Q_+(2n+i) + Q_-(2n+1)) = \psi(Q_+(2n+1)) + \psi(Q_-(2n-1)) \; .$$

It means that $T$ and $I$ are equal in cohomology. So we will have, denoting by $IM(x_1, \cdots, x_N)$ the integrals of motion which can be expressed as $\sum_i P(x_{i+1}, \cdots, x_{i+N})$ [$P$ is a formal series in $x_1 - 1, \cdots, x_N - 1$] :

**Proposition.—**
1) $IM(x_1, \cdots, x_N)$ is a graded linear space, isomorphic to $H^1(a_N) = \bigoplus_{i=0}^{N-1} \mathbf{C}\lambda^i(e + \lambda f)^*$.
2) The natural map $IM(x_1, \cdots, x_N) \to IM(x_1, \cdots, x_{N+1})$ commutes with the natural injection of the corresponding graded spaces.

In other words, we have a "new" integral of motion for each $N$.

Note that by inductive limit the space of all polynomials in $x_1, \cdots, x_N, \cdots$ gets identified with Fun $(\widehat{N}_+/A)$ and we can hope to describe the action of the integrals of motion on this space in a way analogous to [FF]. We can also hope that the elements of $IM(x_1, \cdots, x_N)$ are equivalent to the Faddeev - Volkov integrals of motion ([FV]).

## 8. A geometric interpretation of the screened local quantities.

Let us denote by $\bar{\Sigma}^+$ and $\bar{\Sigma}^-$ the quantities $\sum_{i \geq 0} x_i, \sum_{i \geq 0} y_i$; the action of $\widehat{n}_-$ on the Poisson algebra generated by them, $\mathbf{C}[\bar{\Sigma}^+, \bar{\Sigma}^-, \cdots]$ can be naturally prolonged to an action of $s\ell_2((\lambda))$, on the following way : as $\widehat{n}_-$-module, $\mathbf{C}[\bar{\Sigma}^+, \bar{\Sigma}^-, \cdots]$ can be identified with $\mathbf{C}[\widehat{N}_+]$, endowed with the dressing action ; this action is compatible with the left action of $\widehat{n}_-$, by the mapping $\widehat{N}_+ \to \widehat{G}/\widehat{B}_-$ ; since this mapping is a dense embedding, the left action of $\widehat{g} = s\ell_2((\lambda))$ defines an action of $\widehat{g}$ on $\widehat{N}_+$. Let $\partial_+, \partial_-$ be the operators on $\mathbf{C}[\bar{\Sigma}^+, \bar{\Sigma}^-, \cdots]$, corresponding to the generators of $\widehat{n}_+$. Then $\partial_+$ and $\partial_-$ are derivations, and we have for example $\partial_\varepsilon(\bar{\Sigma}^{\varepsilon'}) = \delta_{\varepsilon \varepsilon'}$, $\varepsilon, \varepsilon' = +$ or $-$.

We can now define the action of $\widehat{g}$ on $\mathbf{C}[x_i, y_i, \bar{\Sigma}^+, \bar{\Sigma}^-, \cdots]_{d \leq i < 0} = \mathbf{C}[x_i, y_i]_{d \leq i < 0} \otimes \mathbf{C}[\bar{\Sigma}^+, \bar{\Sigma}^-, \cdots]$ by $\{\Sigma^\pm, \cdot\}'$ for the $\widehat{n}$-part, and $1 \otimes \partial_\pm$ the $\widehat{n}_+$-part. Recalling that $\{\bar{\Sigma}^\pm, \cdot\}'$ is expressed as $Q_\pm \otimes 1 \pm \frac{1}{2} H \otimes m(\bar{\Sigma}_\pm) + 1 \otimes \{\bar{\Sigma}_\pm, \cdot\}'$ [sect. 2], we see that these formulae indeed define an action of $\widehat{g}$ on $\mathbf{C}[x_i, y_i, \bar{\Sigma}^+, \bar{\Sigma}^-, \cdots]_{d \leq i < 0}$.

Let us interpret this algebra as a function algebra on a homogeneous space of $\widehat{g}$. We have $\mathbf{C}[x_i, y_i]_{d \leq i < 0} \otimes \mathbf{C}[\bar{\Sigma}^+, \bar{\Sigma}^-, \cdots] \simeq \mathbf{C}[\widehat{B}_-/\widehat{B}_- \cap \widehat{B}_-^w \times \widehat{N}_+]$, for a certain Weyl group element $w$ ; on $\widehat{B}_-/\widehat{B}_- \cap \widehat{B}_-^w \times \widehat{N}_+$, the action of $\widehat{n}_+$ is given by the product (0, right translation), and the action of $\widehat{b}_-$ is given by the vector fields $Q_\pm \otimes 1 \pm \frac{1}{2} H \otimes m(\bar{\Sigma}_\pm) + 1 \otimes \{\bar{\Sigma}_\pm, \cdot\}'$. Consider the mapping $\widehat{B}_-/\widehat{B}_- \cap \widehat{B}_-^w \times \widehat{N}_+ \to \widehat{G}/\widehat{B}_- \cap \widehat{B}_-^w$, $(b_-(\widehat{B}_- \cap \widehat{B}_-^w), n_+) \mapsto n_+ b_-(\widehat{B}_- \cap \widehat{B}_-^w)$, and let us show that it is $\widehat{g}$-equivariant. For the part $\widehat{n}_+$, it is clear. The action of $Q_\pm \in \widehat{b}_-$ on the right is $n_+(n_+^{-1} Q_\pm n_+)_+ b_- + n_+(n_+^{-1} Q_\pm n_+)_0 b_- + n_+(n_+^{-1} Q_\pm n_+)_- b_-$ where indices $+, 0$, and $-$ are the projections on the components of $\widehat{g} = \widehat{n}_+ + \mathbf{C} h + \widehat{n}_-$. But, $(n_+^{-1} Q_\pm n_+)_- = Q_\pm$, and $(n_+^{-1} Q_+ n_+)_0 = b_0 \frac{h}{2}, (n_+^{-1} Q_- n_-)_0 = c_1 \frac{h}{2}$, with $n_+ = \exp(b_0 e(0) + c_1 f(1) + + \cdots)$, and $b_0$ and $c_1$ are respectively identified with $\bar{\Sigma}_+$ and $\bar{\Sigma}_-$ by $\mathbf{C}[\bar{\Sigma}_+, \bar{\Sigma}_-, \cdots] \simeq \mathbf{C}[\widehat{N}_+]$ ; this proves the claimed equivariance.

Let us concentrate now on the identification of these spaces as Poisson manifolds. The identification $\mathbf{C}[\bar{\Sigma}^+, \bar{\Sigma}^-, \cdots] \simeq \mathbf{C}[\widehat{N}_+]$ is also an isomorphism of Poisson algebras. Let us identify now the Poisson structure on $\widehat{B}_-/\widehat{B}_- \cap \widehat{B}_-^w$, given by $\mathbf{C}[\widehat{B}_-/\widehat{B}_- \cap \widehat{B}^w] = \mathbf{C}[x_i, y_i]_{-d \leq i \leq 0}$. Recall that the map $\mathbf{C}[\widehat{N}_+] \to \mathbf{C}[x_i, y_i]$ is a Poisson map, and so the map from $(\mathbf{C}^*)^{2d}$ to $\widehat{N}_+ \cap (\widehat{B}_- w \widehat{B}_-)$ [symplectic leaf of $\widehat{N}_+$] is Poisson. Let us consider now the map $\widehat{N}_+ \cap (\widehat{B}_- w \widehat{B}_-) \to \widehat{B}_-/\widehat{B}_- \cap \widehat{B}_-^w$. The group $\widehat{B}_-$ acts in a Lie-Poisson way on both sides ; it means that on the right side the Poisson structure is the structure such that the projection $\widehat{B}_- \to \widehat{B}_-/\widehat{B}_- \cap \widehat{B}_-^w$ is Poisson [such a structure exists since if $s^w = \widehat{b}_- \cap \widehat{b}_-^w$, $\delta(s^w) \subset s^w \cap \widehat{b}_-$], plus some left-invariant bivector. To compute this bivector, we remark that the two maps

$$\widehat{N}_+ \cap (\widehat{B}_- w \widehat{B}_-) \to \widehat{B}_- w \widehat{B}_- \to \widehat{B}_-/\widehat{B}_- \cap \widehat{B}_-^w$$

should also be Poisson, where on the two first spaces the Poisson structures are given by the embeddings in $\widehat{N}_+$ and $\widehat{G}$ respectively. So it is enough to compare the Poisson structures at $w$ in the second space, and at $e(\widehat{B}_- \cap \widehat{B}_-^w)$ in the third one. Denoting by $r$ the element of $\wedge^2 \widehat{g}$ representing the trigonometric $r$-matrix, we find that second bivector is $r - wrw^{-1}$. It means that the Poisson structure on $\widehat{B}_-/\widehat{B}_- \cap \widehat{B}_-^w$ is such that its embedding in $\widehat{G}/\widehat{B}_- \cap \widehat{B}_-^w$, with structure $r^L - (wrw^{-1})^R$ [exponents $L$ and $R$ mean left and right action of $\wedge^2 \widehat{g}$], is Poisson.

Finally, note that the Poisson structure on $\widehat{B}_-/\widehat{B}_- \cap \widehat{B}_-^w \times \widehat{N}_+$ corresponding to that on $\mathbf{C}[x_i, y_i, \bar{\Sigma}_+, \cdots]$ is such that the projections on each factor are Poisson, and $\{f, g\} = (\deg f)(\deg g) f g$, if $f$ and $g$ come respectively from the first and second factor. On the other hand, the map $\widehat{B}_- \times \widehat{N}_+ \to \widehat{G}$, $(b_-, n_+) \mapsto n_+ b_-$, is Poisson, if $\widehat{G}$ has the Poisson structure $r^L - (wrw^{-1})^R$, and $\widehat{B}_- \times \widehat{N}_+$ has the Poisson structure such that first projection on $\widehat{B}_-$, composed with embedding in $\widehat{G}$, with Poisson structure $r^L - (wrw^{-1})^R$, is Poisson, as well as the second projection on $\widehat{N}_+$ with usual Poisson structure, and functions coming from different factors have the same brackets as previously. ($\widehat{G}$ is here the product $\widehat{N}_+ \times \widehat{B}_-$, the factors of this product being completed in the topologies of $\mathbf{C}[[\lambda]]$, resp. $\mathbf{C}[[\lambda^{-1}]]$; it is not a group but has actions of $\widehat{g}$ by left and right translations.)

To summarize, we have :

**Proposition.—** *The morphisms of Poisson algebras $\mathbf{C}[\bar{\Sigma}^\pm, \cdots] \hookrightarrow \mathbf{C}[x_i, y_i, \bar{\Sigma}^\pm, \cdots] \to \mathbf{C}[x_i, y_i]$, (the latter mapping is obtained by factorizing the Poisson ideal generated by $\bar{\Sigma}^+$ and $\bar{\Sigma}^-$) are respectively $\widehat{g}$- and $\widehat{n}_-$-equivariant and are dual to the mappings of Poisson manifolds $\widehat{B}_-/\widehat{B}_- \cap \widehat{B}_-^w \hookrightarrow \widehat{G}/\widehat{B}_- \cap \widehat{B}_-^w \to \widehat{G}/\widehat{B}_-$, where the second manifold has Poisson structure $r^L - (wrw^{-1})^R$ and the third has Poisson structure $r^L$ (where $r \in \wedge^2 \widehat{g}$ is the trigonometric r-matrix). The morphism of Poisson algebras $\mathbf{C}[x_i, y_i] \to \mathbf{C}[x_i, y_i, \bar{\Sigma}^\pm, \cdots]$, defined by $x_i, y_i \mapsto x_i, y_i$, is dual to the projection $\widehat{G}/\widehat{B}_- \cap \widehat{B}_-^w \to \widehat{N}_+ \backslash \widehat{G}/\widehat{B}_- \cap \widehat{B}_-^w \hookleftarrow \widehat{B}_-/\widehat{B}_- \cap \widehat{B}_-^w$.*

Let us now determine to which operations of homogeneous spaces correspond the natural embeddings of algebras of screened local quantities. Let $\alpha < \beta < \gamma$ be three points or the line ; then the embedding $\mathbf{C}[x_i, y_i, \bar{\Sigma}_\pm, \cdots]_{\beta \le i \le \gamma} \hookrightarrow \mathbf{C}[x_i, y_i, \bar{\Sigma}_\pm, \cdots]_{\alpha \le i \le \gamma}$ corresponds to the natural projection $\widehat{G}/\widehat{B}_- \cap \widehat{B}_-^w \to \widehat{G}/\widehat{B}_- \cap \widehat{B}_-^{w'}$ (here $\bar{\Sigma}_+ = \sum_{i > \gamma} x_i$, $\bar{\Sigma}_- = \sum_{i > \gamma} y_i$). Let us pose now $\widehat{\Sigma}_+ = \sum_{i > \beta} x_i$, $\widehat{\Sigma}_- = \sum_{i > \beta} y_i$, and let us consider the embedding $\mathbf{C}[x_i, y_i, \widehat{\Sigma}_\pm, \cdots]_{\alpha \le i \le \beta} \to \mathbf{C}[x_i, y_i, \bar{\Sigma}_\pm, \cdots]_{\alpha \le i \le \gamma}$. It corresponds to the mapping $\widehat{N}_+ \times \mathbf{C}^{2(\gamma - \alpha + 1)} \to \widehat{N}_+ \times \mathbf{C}^{2(\beta - \alpha + 1)}$, $(x_i, y_i, n_+) \mapsto \left( x_i, y_i, n_+ \prod_{i=\gamma}^{\beta+1} \begin{pmatrix} 1 & y_i \\ 0 & 1 \end{pmatrix} \begin{pmatrix} 1 & 0 \\ \lambda x_i & 1 \end{pmatrix} \right)$. The identifications of $\mathbf{C}^{2(\gamma - \alpha + 1)}$ and $\mathbf{C}^{2(\beta - \alpha + 1)}$ with $\widehat{B}_-/\widehat{B}_- \cap \widehat{B}_-^{w_{\gamma - \alpha}}$ and $\widehat{B}_-/\widehat{B}_- \cap \widehat{B}_-^{w_{\beta - \alpha}}$ are $(x_i, y_i) \mapsto$ class of $b_-$, such that $\prod_{i=\gamma}^{\beta+1} \begin{pmatrix} 1 & y_i \\ 0 & 1 \end{pmatrix} \begin{pmatrix} 1 & 0 \\ \lambda x_i & 1 \end{pmatrix} \in b_- w_{\gamma - \beta} \widehat{B}_-$ [resp. same formula with $\beta$ replaced by $\alpha$], so the mapping $\widehat{N}_+ \times \mathbf{C}^{2(\gamma - \alpha + 1)} \to \widehat{G}/\widehat{B}_- \cap \widehat{B}_-^{w_{\gamma - \alpha}}$ is $(n_+, x_i, y_i) \mapsto$ class of $n_+ \prod_{i=\gamma}^{\alpha} \begin{pmatrix} 1 & 0 \\ \lambda y_i & 1 \end{pmatrix} \begin{pmatrix} 1 & x_i \\ 0 & 1 \end{pmatrix} w_{\gamma - \alpha}$. This proves that the initial embedding corresponds to the mapping $\widehat{G}/\widehat{B}_- \cap \widehat{B}_-^{w_{\gamma - \alpha}} \to \widehat{G}/\widehat{B}_- \cap \widehat{B}_-^{w_{\beta - \alpha}}$, class of $g \mapsto$ class of $g w_{\beta - \gamma}$.

The algebra $\mathbf{C}[x_i, y_i, \bar{\Sigma}^\pm_{\geq 0}, \cdots]_{i<0}$ is the union of algebras $\mathbf{C}[x_i, y_i, \bar{\Sigma}^\pm_{\geq 0}, \cdots]_{-N<i<0}$, which is the function algebra on the projective limit of $\cdots \to \widehat{G}/\widehat{B}_- \cap \widehat{B}_-^{w_N} \to \cdots \to \widehat{G}/\widehat{B}_-$. This projective limit is $\widehat{G}/(S^1 \to B)_-$ [where $(S^1 \to B)_-$ is the group corresponding to the Lie algebra $(S^1 \to b)_- = \mathbf{C}[[\lambda^{-1}]] \otimes b]$. The Poisson structure on this space is then $r^L - (w_\infty r w_\infty^{-1})^R$, where $w_\infty r w_\infty^{-1}$ is the $r$-matrix corresponding to the Manin triple $(\widehat{g}, (S^1 \to n) \oplus (S^1 \to \mathbf{C}h)_+, (S^1 \to n_-) \oplus (S^1 \to \mathbf{C}h)_-)$ in notations generalizing the previous one.

Then the embedding $\mathbf{C}[x_i, y_i, \bar{\Sigma}^\pm_{\geq 0}, \cdots]_{i<0} \hookrightarrow \mathbf{C}[x_i, y_i, \bar{\Sigma}^\pm_{\geq N}, \cdots]_{i<N}$ corresponds to the projection $\widehat{G}/(S^1 \to B)_- \to \widehat{G}/(S^1 \to B)_-$, class $(g) \mapsto$ class $(gw_N)$. This projection can be viewed as the composition $\widehat{G}/(S^1 \to B)_- \xrightarrow{\sim} \widehat{G}/w_N^{-1}(S^1 \to B)_- w_N \to \widehat{G}/(S^1 \to B)_-$, where the first map is class$(g) \mapsto$ class$(gw_N)$ and the second is the natural projection. Note that the Poisson structure on the second space, induced by the first map, is $r^L - (w_N^{-1}(w_\infty r w_\infty^{-1}) w_N)^R = r^L - (w_\infty r w_\infty^{-1})^R$.

We obtain :

**Proposition.—** *The inductive limit of algebras $\mathbf{C}[x_i, y_i, \bar{\Sigma}^\pm_{\geq N}, \cdots]_{i<N}$, is identified with functions on $\widehat{G}/(S^1 \to H)_-$, with Poisson structure $r^L - (w_\infty r w_\infty^{-1})^R$, and action of screening operators given by left translations by $\widehat{b}_+$. The inductive limit of algebras $\mathbf{C}[x_i, y_i]_{i<N}$ is identified with functions on $\widehat{B}_-/(S^1 \to H)_-$ ; Poisson structure and injection of this algebra in the latter are given by*

$$\widehat{G}/(S^1 \to H)_- \to \widehat{N}_+ \backslash (S^1 \to H)_- \hookleftarrow \widehat{B}_-/(S^1 \to H)_- .$$

This is because $(S^1 \to H)_- = \cap_N (w_N^{-1}(S^1 \to B)_- w_N)$ [here $H$ is the Cartan subgroup of $B$].

## 9. Commutativity and geometric interpretation of the integrals of motion.

We are now able to give a geometrical description of the Hamiltonian vector fields generated by the integrals found in 6. The action of these vector fields on $\varinjlim \mathbf{C}[x_i, y_i, \bar{\Sigma}^\pm_{\geq N}, \cdots]_{i<N}$ corresponds to vector fields on $\widehat{G}/(S^1 \to H)_-$, commuting with the left action of $\widehat{b}_+$. Let us show that the vector field generated by integral $I_k$ (let us denote it $V_{I_k}$) also commutes with the left action of $\widehat{b}_-$. Indeed, $[V_{I_k}, \partial_\pm]$ should commute with $\{\bar{\Sigma}_\pm, \cdot\}'$. Pose $X_k^\pm = [V_{I_k}, \partial_\pm]$. We compute $X_k^\pm \cdot ($polynomials in $x_i, y_i) = 0$. We deduce that $X_k^\pm$ vanishes on the smallest subalgebra of $\varinjlim \mathbf{C}[x_i, y_i, \bar{\Sigma}^\pm_{\geq N}, \cdots]_{i<N}$ containing the polynomials in $x_i, y_i$, and which is $\{\bar{\Sigma}_\pm, \cdot\}'$ stable ; this algebra is the full algebra, and $X_k^\pm = 0$. So, vector fields $V_{I_k}$ can only be given by right translations by elements of $(S^1 \to \mathbf{C}h)_+$ [$\mathbf{C}h = \mathrm{Lie}(H)$]. In particular, we see that these vector fields commute, and so the integrals of section 6 are in involution.

Remark also that the integral of motion corresponding to $h\lambda^n$ ($n \geq 0$) involves $n$ dots on the line and so should map Fun $(\widehat{G}/w_N^{-1}(S^1 \to B)_- w_N)$ to Fun $(\widehat{G}/(w_N w_n)^{-1} (S^1 \to B)_- w_N w_n)$, it means that it corresponds to the right action of a linear combination of elements $h\lambda^k, 1 \leq k \leq n$.

We conclude :

**Proposition.**— *By the identifications of last proposition, the Hamiltonian vector field corresponding to the integral $I(h\lambda^n)$ found in 6, acts on $\varinjlim \mathbf{C}[x_i, y_i, \bar{\Sigma}^{\pm}_{\geq N}, \cdots]_{i<N}$ and on $\mathbf{C}[x_i, y_i]_{i\in\mathbf{Z}}$ as the right action of a linear combination of elements $h\lambda^k, 1 \leq k \leq n$, on $\widehat{G}/(S^1 \to H)_-$ and on $\widehat{N}_+\backslash\widehat{G}/(S^1 \to H)_-$, respectively. In particular, these integrals are in involution.*

Let us give an explicit form for these vector fields. The identification of $\mathbf{C}[x_i, y_i]_{i\geq 1}$ with $\mathbf{C}[\widehat{B}_-/(S^1 \to B)_-]$ associates to the point $(x_i, y_i)$ the class of the matrix $\begin{pmatrix} 1 & \rho \\ 0 & 1 \end{pmatrix}$, with $\rho = 1/\lambda x_1 + 1/y_1 + 1/\lambda x_2 + \cdots$. Let us describe now the maps $\widehat{B}_-/(S^1 \to B) \xrightarrow{\sim} \widehat{B}_-/w_1^{-1}(S^1 \to B)w_1 \to \widehat{B}_-/(S^1 \to B)$ whose composition is dual to the embedding $\mathbf{C}[y_0, x_i, y_i]_{i\geq 1} \hookrightarrow \mathbf{C}[x_i, y_i]_{i\geq 1}$. The second map is the natural projection, and the first is constructed as follows : to the class of $\begin{pmatrix} 1 & \rho \\ 0 & 1 \end{pmatrix}$ we associate the double class of $b_-$, such that $\begin{pmatrix} 1 & \rho \\ 0 & 1 \end{pmatrix} w_1 \in N_+ b_- w_1^{-1}(S^1 \to B)_- w_1$, with $w_1 = \begin{pmatrix} 0 & -\lambda^{-1} \\ \lambda & 0 \end{pmatrix}$. We have $\begin{pmatrix} 1 & \rho \\ 0 & 1 \end{pmatrix} w_1 \in \begin{pmatrix} 1 & 0 \\ \frac{\lambda}{\rho_1} & 1 \end{pmatrix} \begin{pmatrix} 1 & 0 \\ (\frac{1}{\rho})_{\leq 0} & 1 \end{pmatrix} (S^1 \to B)_-$. At the next step, we multiply by $w_0 = \begin{pmatrix} 0 & -1 \\ 1 & 0 \end{pmatrix}$. Since $\begin{pmatrix} 1 & 0 \\ \sigma & 1 \end{pmatrix} w_0 \in \begin{pmatrix} 1 & \frac{1}{\sigma_0} \\ 0 & 1 \end{pmatrix} \begin{pmatrix} 1 & (\frac{1}{\sigma})_{<0} \\ 0 & 1 \end{pmatrix} (S^1 \to B^{w_0})_-$, we obtain

$$\begin{pmatrix} 1 & \rho \\ 0 & 1 \end{pmatrix} w_1 w_0 \in \begin{pmatrix} 1 & 0 \\ \frac{\lambda}{\rho_1} & 1 \end{pmatrix} \begin{pmatrix} 1 & \frac{1}{\sigma_0} \\ 0 & 1 \end{pmatrix} \begin{pmatrix} 1 & (\frac{1}{\sigma})_{<0} \\ 0 & 1 \end{pmatrix} (S^1 \to B^{w_0})_-,$$

where $\rho = \rho_1 \lambda^{-1} + \cdots$, $\sigma = \sigma_0 + \sigma_1 \lambda^{-1} \cdots = (\frac{1}{\rho})_{\leq 0}$, and indexes $\leq 0$ or $< 0$ mean to take only $\leq 0$ (resp. $< 0$) powers of $\lambda$.

Iterating this procedure we obtain for variables $(x_{-N}, y_{-N}, \cdots)$ the equality

$$\begin{pmatrix} 1 & 1/\lambda x_{-N} + 1/y_{-N} \cdots \\ 0 & 1 \end{pmatrix} \begin{pmatrix} \lambda^{-N} & 0 \\ 0 & \lambda^N \end{pmatrix} = \begin{pmatrix} 1 & 0 \\ \lambda x_{-N} & 1 \end{pmatrix} \begin{pmatrix} 1 & y_{-N} \\ 0 & 1 \end{pmatrix} \cdots \begin{pmatrix} 1 & y_0 \\ 0 & 1 \end{pmatrix} \cdot$$
$$\cdot \begin{pmatrix} 1 & 1/\lambda x_1 + 1/y_1 + \cdots \\ 0 & 1 \end{pmatrix} . \quad \text{element of} \quad (S^1 \to B^{w_0})_-$$

Writing the element of $(S^1 \to B^{w_0})_-$ on the right side $\begin{pmatrix} 1 & 0 \\ b & 1 \end{pmatrix} \begin{pmatrix} a & 0 \\ 0 & a^{-1} \end{pmatrix}$, we find that the first columns of this matrix and of $\lambda^{-N} \begin{pmatrix} 1 & -1/\lambda x_1 + 1/y_1 \cdots \\ 0 & 1 \end{pmatrix} \begin{pmatrix} 1 & -y_0 \\ 0 & 1 \end{pmatrix} \cdot$
$\cdot \begin{pmatrix} 1 & 0 \\ -\lambda x_0 & 1 \end{pmatrix} \cdots \begin{pmatrix} 1 & -y_{-N} \\ 0 & 1 \end{pmatrix} \begin{pmatrix} 1 & 0 \\ -\lambda x_{-N} & 1 \end{pmatrix}$ coincide. We deduce that

$$b = -[(y_0 + 1/\lambda x_0 + 1/y_1 + \cdots + 1/\lambda x_{-N}) + (1/\lambda x_1 + 1/y_1 + \cdots)]^{-1}.$$

In conclusion, we have :

**Proposition.**— *The identification of $\mathbf{C}[x_i, y_i]_{i \in \mathbf{Z}}$ with $\mathbf{C}[\widehat{B}_-/(S^1 \to H)_-]$ associates to the point $(x_i, y_i)$, the class of the matrix*

$$b_- = \begin{pmatrix} 1 & 1/\lambda x_1 + 1/y_1 + \cdots \\ 0 & 1 \end{pmatrix} \begin{pmatrix} 1 & 0 \\ -[(y_0 + 1/\lambda x_0 + 1/y_1 + \cdots) + (1/\lambda x_1 + 1/y_1 + \cdots)]^{-1} & 1 \end{pmatrix}.$$

*The vector fields given by the integrals of sect. 6 are combinations of the flows*

$$\partial_n \bar{b}_- = (b_- \lambda^n h b_-^{-1})_- \bar{b}_-, n \geq 0$$

*[$\bar{b}_-$ is the class of $b_-$ in $\widehat{B}_-/(S^1 \to H)_-$].*

Note that the change of origin point is performed by the sequence of maps

$$\bar{b}_- \in \widehat{B}_-/(S^1 \to H)_- \xrightarrow{\cdot w_1} \widehat{G}/(S^1 \to H)_- \to \widehat{N}_+ \backslash \widehat{G}/(S^1 \to H)_- \hookleftarrow \widehat{B}_-/(S^1 \to H)_-$$

($w_1$ has to be replaced by $w_0$ at the next step). Since $w_0$ and $w_1$ commute with $h\lambda^n, n \geq 0$, these maps commute with the flows, as we could expect.

Observe that expanding $b_-$ in powers of $\lambda^{-1}$, we obtain functions concentrated near the origin ; as the power of $\lambda^{-1}$ increases these functions involve more variables. This reminds the continuous case, where these functions are differential polynomials at the origin, whose degree increases with the power of $\lambda^{-1}$. Note also that the equations obtained have some features of the non-linear Schrödinger equation (intervention of the homogeneous subalgebra).

## 10 Semilocal quantities

The variables treated above where localised near the origin ; the group elements of $\widehat{N}_+ \backslash \widehat{G}/(S^1 \to H)_-$ can be understood as a discrete version of the "monodromy at the origin". To explain this expression recall the situation in continuous case. We have the dressing identity ([DS])

$$\partial + \Lambda + \phi h = n(\partial + \Lambda + \sum_{i=0}^{\infty} u_i \Lambda^{-i}) n^{-1},$$

$\phi = \phi(x)$, $u_i = u_i(x)$ are differential polynomials in $\phi$, $n = n(x)$ a matrix of $\widehat{N}_-$ with coefficients differential polynomials in $\phi$. The monodromy between $a$ and $b$ has the form $n_a e^{(b-a)\Lambda + \sum_{i=0}^{\infty} \int_a^b u_i \Lambda^{-i}} n_b^{-1}$. When $b - a > 0$ and $\lambda \to +\infty$ the asymptotic expansion of this is $\frac{1}{2} e^{(b-a)\sqrt{\lambda}} n_a (1 + \frac{\Lambda}{\sqrt{\lambda}}) e^{\sum_{i=0}^{\infty} \int_a^b u_i \Lambda^{-i}} n_b^{-1}$. For $a = b$, this is identified with $n_a \Lambda n_a^{-1}$, i.e. with the class of $n_a$ in $\widehat{N}_+/A$.

As was shown in [E], the element of $\widehat{N}_+/A$ obtained in this way corresponds to the one provided by the construction of [FF1]. *

---

\* Let us remark that a result of [FF2] follows from this: the fact that the $n$-th KdV flow corresponds to the right action of $a_{-n} = \lambda^n(e + \lambda f)$, indeed, it is shown in [DS] that this flow is $\partial_n(n_a A) = (n_a a_{-n} n_a^{-1})_- n_a A$.

We have also monodromy variables living on the two half lines, $M^a_{-\infty} \sim (1 + \frac{A}{\sqrt{\lambda}})\exp(\sum_{i=0}^{\infty} \int_{-\infty}^{a} u_i \Lambda^{-i}) n_a^{-1}$ and $M_a^{\infty} \sim n_a(1 + \frac{A}{\sqrt{\lambda}})\exp(\sum_{i=0}^{\infty} \int_a^{\infty} u_i \Lambda^{-i})$; let us show how the equivalent variables $M_a^{\pm} = n_a \exp(\pm \sum_{i=0}^{\infty} \int_a^{\pm\infty} u_i \Lambda^{-i})$, can be obtained in terms of screening action.

Let us note that the vector fields $Q^{\pm} = e^{\mp 2\varphi(a)}\{\int_{-\infty}^{\infty} e^{\pm 2\varphi}, \cdot\}$ act on the algebras $\mathbf{C}[\phi^{(i)}(a), \int_a^{\pm\infty} u_i]$. We have then the pairings

$$U\widehat{n}_+ \times \mathbf{C}[\phi^{(i)}(a), \int_a^{\pm\infty} u_i] \to \mathbf{C}[\phi^{(i)}(a), \int_a^{\pm\infty} u_i] \to \mathbf{C}$$

$$(T, P) \mapsto (TP)(\phi, \phi', \cdots, \int_a^{\pm\infty} u_i) \mapsto (TP)(\phi = 0, \phi' = 0, \cdots, \int_a^{\pm\infty} u_i = 0)$$

inducing mappings $\mathbf{C}[\phi^{(i)}(a), \int_a^{\pm\infty} u_i] \to \mathbf{C}[\widehat{N}_-]$.

**Proposition.—** *These mappings are isomorphisms; the projections* $(\phi^{(i)}(a), \int_a^{\pm\infty} u_i) \mapsto (\phi^{(i)}(a))$ *correspond to the natural projection* $\widehat{N}_- \to \widehat{N}_-/A$ ; *the actions of integrals of motions correspond to the restriction to* $\widehat{N}_-$ *of the right action of $a_+$ on* $\widehat{B}_+ \backslash \widehat{G} \hookleftarrow \widehat{N}_-$.

**Proof.** Let $n_0 = \begin{pmatrix} 1 & 0 \\ \phi & 1 \end{pmatrix}$, $n_1 = \begin{pmatrix} 1 & -\frac{\phi}{\lambda} \\ 0 & 1 \end{pmatrix}$; for $i = 0, 1$, let $M^{\pm}(\lambda)_{a;i} = n_i M_a^{\pm}(\lambda)$. Then $M_{0;i}^{\pm}(\lambda)$ (resp. $M_{1;i}^{\pm}(\lambda)$) is a differential polynomial in $-\phi' + \phi^2$ (resp. $\phi' + \phi^2$) and so the action of $Q_+$ (resp. $Q_-$) on it is trivial. So, $Q_+ M_a^{\pm}(\lambda) = -\begin{pmatrix} 0 & 0 \\ 2 & 0 \end{pmatrix} M_a^{\pm}(\lambda)$, $Q_- M_a^{\pm}(\lambda) = \begin{pmatrix} 0 & -\frac{2}{\lambda} \\ 0 & 0 \end{pmatrix} M_a^{\pm}(\lambda)$. This shows that the image in $SL_2(\mathbf{C}[\widehat{N}_-] \otimes \mathbf{C}((\lambda^{-1})))$ of $M_a^{\pm}(\lambda)$ is the canonical matrix of elements of $\mathbf{C}[\widehat{N}_-] \otimes \mathbf{C}((\lambda^{-1}))$. This proves the surjectivity; for the injectivity, we see that elements $\phi^{(i)}(a)$, and $\int_a^{\pm\infty} u_i$ can be obtained by combinations of the coefficients of $\lambda^i$ in matrix elements of $M_a^{\pm}(\lambda)$. The last part follows again from the form of the flows on the dressing operator, shown in [DS] ; $\partial_n M_a^{\pm} = (n_a a_{-n} n_a^{-1})_+ M_a^{\pm} = (M_a^{\pm} a_{-n} (M_a^{\pm})^{-1})_+ M_a^{\pm}$.

The lattice versions of the modules $\mathbf{C}[\phi^{(i)}(a), \int_a^{\pm\infty} u_i]$ are $\mathbf{C}[x_i, y_i, \sum_{k=1}^{\pm\infty} I_i(k)]$; it will be more convenient to analyse first $\mathbf{C}[x_i, y_i, \sum_{k=1}^{\pm\infty} I_i(k), \bar{\Sigma}^{\pm}, \{\bar{\Sigma}^+, \bar{\Sigma}^-\}, \cdots]$ ($I_i(k)$ is the $k$-th conserved density obtained in 6, beginning at point $x_k$). We define an action of $\widehat{g}$ on this algebra as follows: the action on variables $x_i, y_i, \bar{\Sigma}^{\pm}, \cdots$ is unchanged; writing $Q_{\pm} I_i(0) = (T - 1) f_i^{\pm}$ we set $Q_{\pm}(\sum_{k=1}^{+\infty} I_i(k)) = -f_i^{\pm}$ and $Q_{\pm}(\sum_{k=0}^{-\infty} I_i(k)) = f_i^{\pm}$; finally $\partial_{\pm}(\sum_{k=0}^{+\infty} I_i(k)) = H(\sum_{k=0}^{+\infty} I_i(k)) = 0$. By the identification $\mathbf{C}[x_i, y_i]_{i \geq 1} \simeq \mathbf{C}[\widehat{B}_-/(S^1 \to B_-)_-]$, $f_i^+$ and $f_i^-$ are respectively identified with the functions of the class of $\begin{pmatrix} 1 & \rho \\ 0 & 1 \end{pmatrix}$, $0$ and $-\rho_i$ [$\rho = \sum_{i \geq 1} \rho_i \lambda^{-i}$]. Since the embedding of $\mathbf{C}[x_i, y_i]_{i \geq 0}$ in $\mathbf{C}[x_i, y_i, \bar{\Sigma}^{\pm}, \cdots]_{i \in \mathbf{Z}}$ is the natural map $\mathbf{C}[\widehat{B}_-/(S^1 \to$ $B_-)_-] \to \mathbf{C}[\widehat{G}/(S^1 \to H)_-]$, these functions in turn correspond to the functions of the class of $n_+ \begin{pmatrix} 1 & \rho \\ 0 & 1 \end{pmatrix} \begin{pmatrix} 1 & 0 \\ \sigma & 1 \end{pmatrix}$, $0$ and $-\rho_i$ ($n_+ \in \widehat{N}_+$, and $\sigma \in \mathbf{C}[[\lambda^{-1}]]$).

Let us consider now the $\widehat{g}$-module $\mathbf{C}[\widehat{G}/H]$ (the action is by left translations). Write elements of $\widehat{G}$ under the form $n_+ \begin{pmatrix} 1 & \rho \\ 0 & 1 \end{pmatrix} \begin{pmatrix} 1 & 0 \\ \sigma & 1 \end{pmatrix} \begin{pmatrix} e^\alpha & 0 \\ 0 & e^{-\alpha} \end{pmatrix} H$, $\rho, \alpha \in \lambda^{-1}\mathbf{C}[[\lambda^{-1}]]$, $\sigma \in \mathbf{C}[[\lambda^{-1}]]$, $n_+ \in \widehat{N}_+$.

Then if $\alpha = \sum_{i \geq 1} \alpha_i \lambda^{-i}$, the $\alpha_i$ are $\widehat{n}_+$-invariant, $\mathcal{L}_{e\lambda^{-1}}\alpha_i = 0$ and $\mathcal{L}_f \alpha = -\rho$. So we can extend the identification $\mathbf{C}[x_i, y_i, \bar{\Sigma}^\pm, \cdots] \simeq \mathbf{C}[\widehat{G}/(S^1 \to H)_-]$ to

$$\mathbf{C}[x_i, y_i, \bar{\Sigma}^\pm, \cdots, \sum_{k=1}^\infty I_i(k)] \simeq \mathbf{C}[\widehat{G}/H]$$

by $\sum_{k=1}^\infty I_i(k) \mapsto \alpha_i$, and to $\mathbf{C}[x_i, y_i, \bar{\Sigma}^\pm, \cdots, \sum_{k=0}^{-\infty} I_i(k)] \simeq \mathbf{C}[\widehat{G}/H]$, by $\sum_{k=0}^{-\infty} I_i(k) \mapsto -\alpha_i$. The subrings $\mathbf{C}[x_i, y_i, \sum_{k=0}^{\pm\infty} I_i(k)]$ being the intersections of the kernels of $\partial_+$ and $\partial_-$ are then identified to $\mathbf{C}[\widehat{N}_+ \backslash \widehat{G}/H]$.

The elements of $SL_2(\mathbf{C}[x_i, y_i, \sum_{k=0}^{+\infty} I_i(k)] \otimes \mathbf{C}((\lambda^{-1})))$, corresponding to the element $b_-(\lambda)$ of $SL_2(\mathbf{C}\,[\widehat{N}_+\backslash\widehat{G}/H] \otimes \mathbf{C}((\lambda^{-1})))$ (provided by $\widehat{B}_- \to \widehat{N}_+\backslash\widehat{G}$) are then

$$\begin{pmatrix} 1 & 1/\lambda x_1 + 1/y_1 + \cdots \\ 0 & 1 \end{pmatrix} \begin{pmatrix} 1 & 0 \\ -[(y_0 + 1/\lambda x_0 + \cdots) + (1/\lambda x_1 \cdots)]^{-1} & 1 \end{pmatrix} e^{\sum_{i \geq 1} \lambda^{-i} h \sum_{k \geq 1} I_i(k)}$$

and

$$\begin{pmatrix} 1 & 1/\lambda x_1 + 1/y_1 + \cdots \\ 0 & 1 \end{pmatrix} \begin{pmatrix} 1 & 0 \\ -[(y_0 + 1/\lambda x_0 + \cdots) + (1/\lambda x_1 \cdots)]^{-1} & 1 \end{pmatrix} e^{-\sum_{i \geq 1} \lambda^{-i} h \sum_{k \leq 0} I_i(k)}.$$

Recalling the identities

$$\exp(\sum_{i \geq 1} \lambda^{-i} \sum_{k \geq 1} I_i(k)) = \frac{1/\lambda x_1 + 1/y_1 + \cdots}{1/\lambda x_1} \frac{1/y_1 + 1/\lambda x_2 + \cdots}{1/y_1} \cdots$$

and

$$\exp(-\sum_{i \geq 1} \lambda^{-i} \sum_{k \geq 0} I_i(k)) = \frac{1/y_0}{1/y_0 + 1/\lambda x_1 + \cdots} \frac{1/\lambda x_0}{1/\lambda x_0 + 1/y_0 + \cdots} \frac{1/y_{-1}}{1/y_{-1} + 1/\lambda x_0 + \cdots} \cdots$$

we obtain the form of these matrices in terms of variables $x_i$ and $y_i$.

Let us study now the homogeneous spaces interpretation of lattice translation. From the equalities

$$\begin{pmatrix} 1 & a \\ 0 & 1 \end{pmatrix} \begin{pmatrix} 1 & 0 \\ b & 1 \end{pmatrix} \begin{pmatrix} t & 0 \\ 0 & t^{-1} \end{pmatrix} \begin{pmatrix} 0 & -\lambda^{-1} \\ \lambda & 0 \end{pmatrix} = \begin{pmatrix} 1 & 0 \\ \lambda & 1 \end{pmatrix} \begin{pmatrix} 1 & 0 \\ \frac{1}{a} - \frac{\lambda}{a_1} & 1 \end{pmatrix} \cdot$$
$$\cdot \begin{pmatrix} 1 & -a(1+ab) \\ 0 & 1 \end{pmatrix} \begin{pmatrix} a\lambda t^{-1} & 0 \\ 0 & t/(a\lambda) \end{pmatrix}$$

and

$$\begin{pmatrix} 1 & 0 \\ b & 1 \end{pmatrix} \begin{pmatrix} 1 & a \\ 0 & 1 \end{pmatrix} \begin{pmatrix} t & 0 \\ 0 & t^{-1} \end{pmatrix} \begin{pmatrix} 0 & -1 \\ 1 & 0 \end{pmatrix} = \begin{pmatrix} 1 & \frac{1}{b_0} \\ 0 & 1 \end{pmatrix} \begin{pmatrix} 1 & \frac{1}{b} - \frac{1}{b_0} \\ 0 & 1 \end{pmatrix} \cdot$$
$$\cdot \begin{pmatrix} 1 & 0 \\ -b(1+ab) & 1 \end{pmatrix} \begin{pmatrix} -1/(bt) & 0 \\ 0 & -bt \end{pmatrix},$$

with $a \in \lambda^{-1}\mathbf{C}[[\lambda^{-1}]]$, $b \in \mathbf{C}[[\lambda^{-1}]]$ and $t \in \mathbf{C}[[\lambda^{-1}]]^\times$, follows that right multiplication by affine Weyl group elements in $\widehat{N}_+\backslash\widehat{G}/H$ transform the above matrices into the matrices with shifted arguments $x_i$ and $y_i$. This refines the result obtained previously about translations.

Let us describe now the elements of $SL_2(\mathbf{C}[x_i, y_i, \sum_{k=0}^{\pm\infty} I_i(k), \bar{\Sigma}^\pm, \cdots] \otimes \mathbf{C}((\lambda, \lambda^{-1})))$ corresponding to the element $g(\lambda)$ of $SL_2(\mathbf{C}[\widehat{G}/H] \otimes \mathbf{C}((\lambda, \lambda^{-1})))$ provided by the projection $\widehat{G} \to \widehat{G}/H$. The projection $\widehat{G}/H \to \widehat{G}/(S^1 \to H)_-$ sends it to

$$\prod_{i=\infty}^{1} \begin{pmatrix} 1 & -y_i \\ 0 & 1 \end{pmatrix} \begin{pmatrix} 1 & 0 \\ -\lambda x_i & 0 \end{pmatrix} \cdot \begin{pmatrix} 1 & 1/\lambda x_1 \cdots \\ 0 & 1 \end{pmatrix} \begin{pmatrix} 1 & 0 \\ -[y_0 \cdots]^{-1} & 1 \end{pmatrix}$$

according to 8, and the projection $\widehat{G}/H \to \widehat{N}_+\backslash\widehat{G}/H$ to

$$\begin{pmatrix} 1 & 1/\lambda x_1 + \cdots \\ 0 & 1 \end{pmatrix} \begin{pmatrix} 1 & 0 \\ -[y_0 \cdots]^{-1} & 1 \end{pmatrix} e^{\sum_{i\geq 1} \lambda^{-i} h \sum_{k\geq 1} I_i(k)}$$

$[\begin{pmatrix} 1 & 1/\lambda x_1 + \cdots \\ 0 & 1 \end{pmatrix} \begin{pmatrix} 1 & 0 \\ -[y_0 \cdots]^{-1} & 1 \end{pmatrix} \exp(-\sum_{i\geq 1} \lambda^{-i} h \sum_{k\leq 0} I_i(k))$, resp.]. So these matrices are

$$\prod_{i=+\infty}^{1} \begin{pmatrix} 1 & -y_i \\ 0 & 1 \end{pmatrix} \begin{pmatrix} 1 & 0 \\ -\lambda x_i & 0 \end{pmatrix} \cdot \begin{pmatrix} 1 & 1/\lambda x_1 \cdots \\ 0 & 1 \end{pmatrix} \begin{pmatrix} 1 & 0 \\ -[y_0 \cdots]^{-1} & 1 \end{pmatrix} e^{\sum_{i\geq 1} \lambda^{-i} h \sum_{k\geq 1} I_i(k)}$$

and

$$\prod_{i=+\infty}^{1} \begin{pmatrix} 1 & -y_i \\ 0 & 1 \end{pmatrix} \begin{pmatrix} 1 & 0 \\ -\lambda x_i & 0 \end{pmatrix} \cdot \begin{pmatrix} 1 & 1/\lambda x_1 \cdots \\ 0 & 1 \end{pmatrix} \begin{pmatrix} 1 & 0 \\ -[y_0 \cdots]^{-1} & 1 \end{pmatrix} e^{-\sum_{i\geq 1} \lambda^{-i} h \sum_{k\leq 0} I_i(k)}.$$

Let us pass to the Poisson structures on $\widehat{G}/H$ and $\widehat{N}_+\backslash\widehat{G}/H$ induced by these mappings. Note first that $\widehat{g}$ acts in a Lie-Poisson way not only on $\mathbf{C}[x_i, y_i, \bar{\Sigma}^\pm, \cdots]$, but also on $\mathbf{C}[x_i, y_i, \bar{\Sigma}^\pm, \cdots, \sum_{k=0}^{\pm\infty} I_i(k)]$. Recall that this means that $X\{f,g\} = \sum X^{(1)} f X^{(2)} g + \{Xf, g\} + \{f, Xg\}$ for $X \in \widehat{g}$, $f, g$ in this function algebra, with $\delta X = \sum X^{(1)} \otimes X^{(2)}$ the cobracket of $\widehat{g}$. We can check it replacing $\sum_{k=0}^{\pm\infty} I_i(k)$ by $(1 - T^{\pm N}) \sum_{k=0}^{\pm\infty} I_i(k)$ and letting $N$ to $\pm\infty$, using that $T$ is a $\widehat{g}$-module map.

We thus obtain that the Poisson structure on $\widehat{G}/H$ is of the form $r^L - r'^R$; moreover, the projection on $\widehat{G}/(S^1 \to H)_-$ with structure $r^L - (w_\infty r w_\infty^{-1})^R$ is Poisson, so $r' = w_\infty r w_\infty^{-1} + r_0$, $r_0 \in (S^1 \to h)_- \wedge \widehat{g}$.

The space $\widehat{N}_+\backslash \widehat{G}/H$ has a Poisson structure corresponding to the identification of its function algebra with $\mathbf{C}[x_i, y_i, \sum_{k=0}^{\pm\infty} I_i(k)]$ ; it is such that the projection $\widehat{G}/H \to \widehat{N}_+\backslash \widehat{G}/H$ is Poisson, so it is given by the bivector $-r'^R$. The right multiplication by $w_2 = \begin{pmatrix} \lambda & 0 \\ 0 & \lambda^{-1} \end{pmatrix}$ is an automorphism of this Poisson manifold, since it corresponds to the translation $x_i \mapsto x_{i+1}, y_i \mapsto y_{i+1}$. So we have $r'^{w_2} = r'$, and so $r_0 \in (S^1 \to h)_- \wedge (S^1 \to h)$.

Let us show now that $r_0 \in (S^1 \to h)_+ \wedge (S^1 \to h)_-$. Write the element of $SL_2(\mathbf{C}[x_i, y_i, \sum_{k\geq 0} I_i(k)] \otimes \mathbf{C}((\lambda^{-1})))$ under the form $b_- = \begin{pmatrix} a & c \\ b & d \end{pmatrix}$ ; then $d = \exp(-\sum_{i\geq 1} \lambda^{-i} \sum_{k\geq 0} I_i(h))$. The Poisson brackets $\{I_i(k), I_j(l)\}$ are polynomials in $x_i^{-1}, y_i^{-1}$ without constant terms (since the $I_i(k)$ themselves are polynomials in $x_i^{-1}, y_i^{-1}$). So we should have $\{d(\lambda), d(\mu)\}_{|d_i=\delta_{i_0}} = 0$. On the other hand, if $x \in \widehat{n}_+$, $R(x)b_- = (b_-xb_-^{-1})_- b_-$ vanishes at the origin of $\widehat{B}_-$. So the value of $r'$ at the origin is the projection of $r_0$ in $\wedge^2(S^1 \to h)_-$ along $(S^1 \to h)_- \wedge (S^1 \to h)_+$. Writing this projection $\sum r_{\alpha\beta} h_\alpha \wedge h_\beta$, we find $\{d(\lambda), d(\mu)\}_{\text{origin}} = \sum r_{\alpha\beta} \lambda^\alpha \mu^\beta$ ; so this projection is zero.

Let us try now to determine $r_0$. Elements of $\widehat{G}/H$ being written $gH = n_+n_-H$, with $n_- = \begin{pmatrix} a & c \\ b & d \end{pmatrix}$ and $n_+ = \begin{pmatrix} 1 + \cdots & c_0 + \cdots \\ \lambda b_1 + \cdots & 1 + \cdots \end{pmatrix}$ we should have $\{b_1, d\} = d\rho = c$, $\{c_0, d\} = 0$ by the identifications $b_1 = -\bar{\Sigma}^+, c_0 = -\bar{\Sigma}^-$ and $d = e^{\sum_{i\geq 1} \lambda^{-i} \sum_{k\geq 1} I_i(k)}$ (because on $\sum_{k\geq 1} I_i(k)$, the Poisson brackets with $b_1$ and $c_0$ coincide with the actions of $Q_\pm$). After computations

$$r^L(b_1 \otimes d) = 0 ,$$

$$\left(\sum_{i\in\mathbf{Z}} e\lambda^i \otimes f\lambda^{-i} - f\lambda^i \otimes e\lambda^{-i}\right)^R (b_1 \otimes d) = -\lambda^{-1}bd^2,$$

$$\left(\sum_{i>0} h\lambda^i \otimes h\lambda^{-i} - h\lambda^{-i} \otimes h\lambda^i\right)^R (b_1 \otimes d) = -2\lambda^{-1}bd^2$$

and

$$r^L(c_0 \otimes d) = c ,$$

$$\left(\sum_{i\in\mathbf{Z}} e\lambda^i \otimes f\lambda^{-i} - f\lambda^i \otimes e\lambda^{-i}\right)^R (c_0 \otimes d) = bc^2 ,$$

$$\left(\sum_{i>0} h\lambda^i \otimes h\lambda^{-i} - h\lambda^{-i} \otimes h\lambda^i\right) (c_0 \otimes d) = 2acd ,$$

so
$$r^L - \left(\sum_{i \in \mathbf{Z}}(e\lambda^i \otimes f\lambda^{-i} - f\lambda^i \otimes e\lambda^{-i}) - \frac{1}{2}\sum_{i>0}(h\lambda^i \otimes h\lambda^{-i} - h\lambda^{-i} \otimes h\lambda^i)\right)^R$$
$$= r^L - (w_\infty r w_\infty^{-1})^R$$
give the right brackets $\{b_1, d\}$ and $\{c_0, d\}$.

Let us show that this structure is the only possible on $\widehat{G}/H$ : we know that any other structure differs from this one by the addition of some $A^R$, with $A \in (S^1 \to h)_+ \wedge (S^1 \to h)_-$. On the other hand, functions of $n_+$ should commute with $d$ in for the structure defined by $A$ because we have $\{b_1, d\}_A = \{c_0, d\}_A = 0$ and the Poisson algebra generated by $b_1$ and $c_0$ is the set of functions of $n_+$. But

$$(\sum A_{ij} h\lambda^i \otimes h\mu^j)^R(n_+ \otimes d) = \sum_{i>0, j<0} A_{ij} n_+(n_- h\lambda^i n_-^{-1})_+ \mu^j d(\mu) ,$$

so for any $j$, $\sum A_{ij}(n_- h\lambda^i n_-^{-1})_+ = 0$, so $A = 0$.

Let us consider again $g = n_+ \begin{pmatrix} 1 & \rho \\ 0 & 1 \end{pmatrix} \begin{pmatrix} 1 & 0 \\ \sigma & 1 \end{pmatrix} e^{\sum_{i \geq 1} \lambda^{-i} h \sum_{k \geq 0} \hat{I}_i(k)}$. If $w_0 = \begin{pmatrix} 0 & -\lambda^{-1} \\ \lambda & 0 \end{pmatrix}$, $gw_0 = n'_+ \begin{pmatrix} 1 & 0 \\ \sigma' & 1 \end{pmatrix} \begin{pmatrix} 1 & \rho' \\ 0 & 1 \end{pmatrix} e^{-\sum_{i \geq 1} \lambda^{-i} h \delta_i}$, with $n'_+ = n_+ \begin{pmatrix} 1 & -y_0 \\ 0 & 1 \end{pmatrix}$ and $\sigma' = -1/y_0 + 1/\lambda x_0 + 1/y_{-1}, \cdots$, $\rho' = [\lambda x_1 + 1/y_1 + \cdots + (1/y_0 + 1/\lambda x_0 \cdots)]^{-1}$. If $\delta = \sum_{i \geq 1} \delta_i \lambda^{-i}$, $\mathcal{L}_{e\lambda^{-1}}\delta = -\lambda^{-1}\sigma'$, $\mathcal{L}_f \delta = 0$. On the other hand if $\hat{I}_i(k)$ is the integral of motion obtained in 6, ending at $y_k$, we have $Q_-(\sum_{k \leq 0} \hat{I}_i(k)) = 0$ and $Q_+(\sum_{k \leq 0} \hat{I}_i(k)) = g_i^+$, which is identified with $\sigma_{i-1}$. This allows to identify $\delta$ with $-\sum_{i \geq 1} \lambda^{-i} \sum_{k > 0} \hat{I}_i(k)$ in the case of the module generated by $\sum_{k \geq 0} I_i(k)$, and with $\sum_{i \geq 1} \lambda^{-i} \sum_{k \leq 0} \hat{I}_i(k)$ in the other case. So, $gw_0 e^{-\sum_{i \geq 1} \lambda^{-i} h I_i} = n'_+ \begin{pmatrix} 1 & 0 \\ \sigma' & 1 \end{pmatrix} \begin{pmatrix} 1 & \rho' \\ 0 & 1 \end{pmatrix}$ $\cdot e^{-\sum_{i \geq 1} \lambda^{-i} h \sum_{k \leq 0} \hat{I}_i(k)} = g'$. Let us determine the Poisson brackets of $g'$.

The Poisson bracket $\{I_i, g\}$ is a right translation $ga_i$, $a_i \in (S^1 \to h)_+$ [we know from sect. 9 that it has the form $\{I_i, g\} = g(a_i + (S_1 \to h)_-)$, and the brackets $\{g \otimes, ge^{-\sum_{i \geq 1} I_i \lambda^{-i} h}\}$ have the form $r^L - (w_\infty r w_\infty^{-1})^R - \sum_i (a_i + (S^1 \to h)_-) \otimes \lambda^{-i} h)^R$ ; the action of $\hat{g}$ on $\{g \otimes, ge^{-\sum_{i \geq 1} I_i, \lambda^{-i} h}\}$ is again Poisson-Lie, so the vectors of $(S^1 \to h)_-$ are constant ; repeating the reasoning above, we see that the bivector has no $(S^1 \to h)_- \otimes (S^1 \to h)_-$ components, and $\{I_i, g\} = ga_i$.] Then the Poisson brackets for $gw_0$ are given by $r^L - (w_0 w_\infty r w_\infty^{-1} w_0^{-1})^R$, and on $gw_0 e^{-\sum_{i \geq 1} \lambda^{-i} h I_i}$ by $r^L - (w_0 w_\infty r w_\infty^{-1} w_0^{-1})^R - \sum_{i \geq 1}(\lambda^{-i} h \otimes a_i - a_i \otimes \lambda^{-i} h)^R$.

Let us determine now the Poisson brackets of $g'$ using the writing $g' = n' \begin{pmatrix} 1 & 0 \\ \sigma' & 1 \end{pmatrix}$ $\cdot \begin{pmatrix} 1 & \rho' \\ 0 & 1 \end{pmatrix} e^{-\sum_{i \geq 1} \lambda^{-i} h \sum_{k \leq 0} \hat{I}_i(k)}$. Pose $n' = \begin{pmatrix} 1 + \cdots & c'_0 + \cdots \\ \lambda b'_1 + \cdots & 1 + \cdots \end{pmatrix}$, and $\begin{pmatrix} 1 & 0 \\ \sigma' & 1 \end{pmatrix} \cdot$

$$\cdot \begin{pmatrix} 1 & \rho' \\ 0 & 1 \end{pmatrix} e^{-\sum_{i \geq 1} \lambda^{-i} h \sum_{k \leq 0} \hat{I}_i(k)} = \begin{pmatrix} \bar{a} & \bar{c} \\ \bar{b} & \bar{d} \end{pmatrix}.$$ Then $\{b'_1, \bar{a}\} = -Q_+(\bar{a}) = \lambda^{-1}\sigma'\bar{a} = \lambda^{-1}\bar{b}$, and $\{c'_0, \bar{a}\} = -Q_-(\bar{a}) = 0$. But

$$r^L(c'_0 \otimes \bar{a}) = 0,$$

$$\left(\sum_{i \in \mathbf{Z}} e\lambda^i \otimes f\lambda^{-i} - f\lambda^i \otimes e\lambda^{-i}\right)^R (c'_0 \otimes \bar{a}) = \bar{c}\bar{a}^2,$$

$$\left(\sum_{i > 0} h\lambda^i \otimes h\lambda^{-i} - h\lambda^{-i} \otimes \lambda^i\right)^R (c'_0 \otimes \bar{a}) = -2\bar{c}\bar{a}^2,$$

and

$$r^L(b'_1 \otimes \bar{a}) = \lambda^{-1}\bar{b}$$

$$\left(\sum_{i \in \mathbf{Z}} e\lambda^i \otimes f\lambda^{-i} - f\lambda^i \otimes e\lambda^{-i}\right)^R (b'_1 \otimes \bar{a}) = -\lambda^{-1}\bar{c}\bar{b}^2$$

$$\left(\sum_{i > 0} h\lambda^i \otimes h\lambda^{-i} - h\lambda^{-i} \otimes h\lambda^i\right)^R (b'_1 \otimes \bar{a}) = 2\lambda^{-1}\bar{a}\bar{b}\bar{d}$$

so

$$r^L - (w_0 w_\infty r w_\infty^{-1} w_0^{-1})^R - (\sum_{i > 0} \lambda^{-i} h \otimes a_i - a_i \otimes \lambda^{-i} h)^R$$

$$= r^L - \sum_{i \in \mathbf{Z}}(e\lambda^i \otimes f\lambda^{-i} - f\lambda^i \otimes e\lambda^{-i})^R - \frac{1}{2}\sum_{i > 0}(h\lambda^i \otimes h\lambda^{-i} - h\lambda^{-i} \otimes h\lambda^i)^R$$

This shows (since $w_\infty r w_\infty^{-1}$ is $w_0$-invariant) that $a_i = -h\lambda^i$.

**Theorem.—** *For each point of the lattice, there is a natural $\widehat{g}$-equivariant mapping of the manifold with coordinates $(x_i, y_i, \bar{\Sigma}^\pm, \cdots, \sum_{k=0}^{\pm\infty} I_i(k))$ to $\widehat{G}/H$ ; the mapping corresponding to $x_i$ is $(x_i, y_i, \cdots) \mapsto \prod_{j=+\infty}^{i} \begin{pmatrix} 1 & -y_{i-1} \\ 0 & 1 \end{pmatrix} \begin{pmatrix} 1 & 0 \\ -\lambda x_j & 1 \end{pmatrix} \cdot$
$\begin{pmatrix} 1 & \rho_i \\ 0 & 1 \end{pmatrix} \begin{pmatrix} 1 & 0 \\ \sigma_i & 1 \end{pmatrix} e^{\sum_{i \geq 1} h\lambda^{-i} \sum_{k \geq 0(k < 0)} \pm I_i(k)}$, for point $y_i$ it is $(x_i, y_i, \cdots) \mapsto \prod_{j=+\infty}^{i}$
$\begin{pmatrix} 1 & 0 \\ -\lambda x_j & 1 \end{pmatrix} \begin{pmatrix} 1 & -y_j \\ 0 & 1 \end{pmatrix} \begin{pmatrix} 1 & 0 \\ \sigma'_i & 1 \end{pmatrix} \begin{pmatrix} 1 & \rho'_i \\ 0 & 1 \end{pmatrix} e^{\sum_{i \geq 1} h\lambda^{-i} \sum_{k \leq 0(k > 0)} \widehat{I}_i(k)}$, with $\rho_i = 1/\lambda x_i + 1/y_i + \cdots$, $\sigma_i = -[y_{i-1} + 1/\lambda x_{i-1} \cdots + (1/\lambda x_i + 1/y_i \cdots)]^{-1}$, $\rho'_i = [\lambda x_{i+1} + 1/y_{i+1} + \cdots + (1/y_i + 1/\lambda x_i + \cdots)]^{-1}$, $\sigma'_i = -1/y_i + 1/\lambda x_i + 1/y_{i-1} \cdots$. These mappings are Poisson if we endow $\widehat{G}/H$ with the structure $r^L - (w_\infty r w_\infty^{-1})^R$ [resp. $r^L - (w_{-\infty} r w_{-\infty}^{-1})^R$], where $r$ is the trigonometric $r$-matrix of $\widehat{g}$ and superscripts $L$ and $R$ denote the bivector field generated by left and right action of a given element. The Hamiltonian flow generated by the $i$-th integral of motion $\sum_{k \in \mathbf{Z}} I_i(k)$ corresponds by this mapping to* right translation by $-h\lambda^i$. The sine-Gordon flow corresponds to left translation by $e + \lambda f$.

The passage from the mapping corresponding to point $x_i$ to $y_i$ (resp. from $y_i$ to $x_{i+1}$ is realised by right multiplication by the affine Weyl group element $w_0$ (resp. $w_1$).

Here $w_{-\infty} r w_{-\infty}^{-1}$ corresponds to the Manin triple $(\widehat{g}, (S^1 \to \mathbf{C}h)_+ \oplus (S^1 \to n_-), (S^1 \to \mathbf{C}h)_- \oplus (S^1 \to n))$.

It is possible to define, at the matrix level "higher sine-Gordon flows" by the left translations by other elements of the principal subalgebra, $e\lambda^i + f\lambda^{i+1}$ ($i \geq 1$), commuting to the sine-Gordon and the "mKdV" ones (generated by the integrals of motion). We can think that these flows correspond to some differential equations on variables $x_k$ and $y_k$, which would become more and more non-local as $i$ increases.

**Acknowledgements.** We would like to thank E. Frenkel and V. Lyubashenko for useful discussions, and Mmes Dezetter, François and Harmide for typing this text.

B.E.: Centre de mathématiques
URA 169 du CNRS
Ecole Polytechnique
91128 Palaiseau Cedex
France

B.F.: Landau Inst. for Theor. Physics
Kosygina 2, GSP-1
117940 Moscow V-334
Russia